\documentclass[prb,twocolumn,showpacs,superscriptaddress,amsmath,amssymb,showkeys]{revtex4-2}
\usepackage{natbib}
\usepackage{graphicx}         
\usepackage{dcolumn}          
\usepackage{bm}               
\usepackage{lipsum}
\usepackage{upgreek}
\usepackage{booktabs} 				
\usepackage{tabularx}
\usepackage{array}
\usepackage[colorlinks=true,urlcolor=blue,breaklinks=true]{hyperref}
\usepackage{xcolor}
\usepackage{colortbl}
\usepackage{multirow}
\usepackage{afterpage}
\usepackage[version=3]{mhchem}
\usepackage{mathtools}
\usepackage{esint}
\usepackage{slashbox}
\usepackage[indent=12pt,skip=0pt]{parskip}
\definecolor{green2}{RGB}{15, 117, 19}

\definecolor{Orange}{RGB}{255, 165, 19}

\begin{document}

\newcommand*{\citen}[1]{
  \begingroup
    \romannumeral-`\x
    \setcitestyle{numbers}
    \cite{#1}
  \endgroup   
}

\title{Flux-periodic supercurrent oscillations in an Aharonov--Bohm--type nanowire Josephson junction}

\author{Patrick Zellekens}
\affiliation{RIKEN Center for Emergent Matter Science, 351-0198 Wako, Japan}
\email{p.zellekens@mailbox.org}

\author{Russell S. Deacon} \affiliation{RIKEN Center for Emergent Matter Science, 351-0198 Wako, Japan} \affiliation{Advanced Device Laboratory, RIKEN, 351-0198 Wako, Japan} 

\author{Farah Basaric} 
\affiliation{Peter Gr\"unberg Institut 9, Forschungszentrum J\"ulich, 52425 J\"ulich, Germany} \affiliation{JARA-Fundamentals of Future Information Technology, J\"ulich-Aachen Research Alliance, Forschungszentrum J\"ulich and RWTH Aachen University, 52425 J\"ulich, Germany}

\author{Raghavendra Juluri} \affiliation{Department of Physics, University of Warwick, Coventry CV4 7AL, UK}

\author{Michael D. Randle}
\affiliation{Advanced Device Laboratory, RIKEN, 351-0198 Wako, Japan}

\author{Benjamin Bennemann} 
\affiliation{Peter Gr\"unberg Institut 10, Forschungszentrum J\"ulich, 52425 J\"ulich, Germany} \affiliation{JARA-Fundamentals of Future Information Technology, J\"ulich-Aachen Research Alliance, Forschungszentrum J\"ulich and RWTH Aachen University, 52425 J\"ulich, Germany}

\author{Christoph Krause} 
\affiliation{Peter Gr\"unberg Institut 10, Forschungszentrum J\"ulich, 52425 J\"ulich, Germany} \affiliation{JARA-Fundamentals of Future Information Technology, J\"ulich-Aachen Research Alliance, Forschungszentrum J\"ulich and RWTH Aachen University, 52425 J\"ulich, Germany}

\author{Erik Zimmermann} 
\affiliation{Peter Gr\"unberg Institut 9, Forschungszentrum J\"ulich, 52425 J\"ulich, Germany} \affiliation{JARA-Fundamentals of Future Information Technology, J\"ulich-Aachen Research Alliance, Forschungszentrum J\"ulich and RWTH Aachen University, 52425 J\"ulich, Germany}

\author{Ana M. Sanchez} \affiliation{Department of Physics, University of Warwick, Coventry CV4 7AL, UK}

\author{Detlev Gr\"utzmacher} \affiliation{Peter Gr\"unberg Institut 9, Forschungszentrum J\"ulich, 52425 J\"ulich, Germany}\affiliation{JARA-Fundamentals of Future Information Technology, J\"ulich-Aachen Research Alliance, Forschungszentrum J\"ulich and RWTH Aachen University, 52425 J\"ulich, Germany}

\author{Alexander Pawlis} \affiliation{Peter Gr\"unberg Institut 10, Forschungszentrum J\"ulich, 52425 J\"ulich, Germany} \affiliation{JARA-Fundamentals of Future Information Technology, J\"ulich-Aachen Research Alliance, Forschungszentrum J\"ulich and RWTH Aachen University, 52425 J\"ulich, Germany}

\author{Koji Ishibashi} \affiliation{RIKEN Center for Emergent Matter Science, 351-0198 Wako, Japan} \affiliation{Advanced Device Laboratory, RIKEN, 351-0198 Wako, Japan}

\author{Thomas~Sch\"apers}\affiliation{Peter Gr\"unberg Institut 9, Forschungszentrum J\"ulich, 52425 J\"ulich, Germany} \affiliation{JARA-Fundamentals of Future Information Technology, J\"ulich-Aachen Research Alliance, Forschungszentrum J\"ulich and RWTH Aachen University, 52425 J\"ulich, Germany} 

\keywords{InAs nanowire, selective-area growth, molecular beam
epitaxy, Josephson junctions, mesoscopic transport}
\date{\today}

\begin{abstract}
Phase winding effects in hollow semiconductor nanowires with superconducting shells have been proposed as a route to engineer topological superconducting states. We investigate GaAs/InAs core/shell nanowires with half-shells of epitaxial aluminium as a potential platform for such devices, where the thin InAs shell confines the electron wave function around the GaAs core. With normal contacts we observed pronounced $h/e$ flux periodic oscillations in the magnetoconductance, indicating the presence of a tubular conductive channel in the InAs shell. Conversely, the switching current in Josephson junctions oscillates with approximately half that period, i.e. $h/2e$, indicating transport via Andreev transport processes in the junction enclosing threading magnetic flux. On these structures, we systematically studied the gate-, field-, and temperature-dependent evolution of the supercurrent. Results indicate that Andreev transport processes can occur about the wire circumference indicating full proximitization of the InAs shell from the half-shell superconducting contacts.
\end{abstract}

\maketitle

\section{Introduction}

Josephson junctions, at their core, are two superconductors separated by a thin insulating barrier or a weak link. They allow the dissipationless and coherent transfer of bosonic quasiparticles, i.e. Cooper pairs. The resulting supercurrent flows without the application of an external voltage and is only driven by the phase difference between the macroscopic wavefunctions in the superconducting electrodes, i.e. the Josephson effect.

Realizations of such devices with a semiconductor nanowire (NW) as the weak link have generated significant attention for their potential in the realization of superconducting quantum circuits, particularly within the realm of quantum computing.  The appeal of these junctions lies in their multifaceted advantages. Firstly, the incorporation of a semiconducting weak link enables precise control of junction parameters through field-effect manipulation. This attribute permits dynamic adjustments of the quantum circuit's properties, making it a flexible platform for quantum information processing. Secondly, the unique combination of a large Fermi wavelength and a small nanowire diameter enables the design of junctions wherein only a limited subset of quantum states, known as Andreev bound states, play a pivotal role in carrying the Josephson supercurrent \cite{Tosi2019,Metzger2021,Zellekens2022}. This selective modulation facilitates the use of paired Andreev bound states as a foundational element for the creation of quantum bits (qubits)\cite{Zazunov2003,Zazunov2005,Woerkom2017,Hays2021,Cerrillo2021}, which are the fundamental units of quantum information processing.

Moreover, semiconductor nanowire hybrid structures play a central role in the quest for realizing topological qubits \cite{Sarma2015,Oreg2010,Lutchyn2010}, a cutting-edge approach to fault-tolerant quantum computing beyond the current NISQ (noisy-intermediate-scale-quantum) methodology. Here, qubits are constructed based on the unique properties of topological states of matter, which offer inherent protection against certain types of quantum errors, thus rising the fidelity of the whole system. A key element in this endeavor is the search for parafermionic excitations, e.g. Majorana zero modes (MZMs) \cite{Sarma2015,Kitaev2003}. The latter are a particular type of quasiparticle excitation that behaves as its own antiparticle, i.e. with non-abelian characteristics. Thus, they are decoupled from the conventional charge carrier space, which makes them highly robust against external disturbances. 
\begin{figure*}[!t]
	\centering
\includegraphics[width=0.9\linewidth]{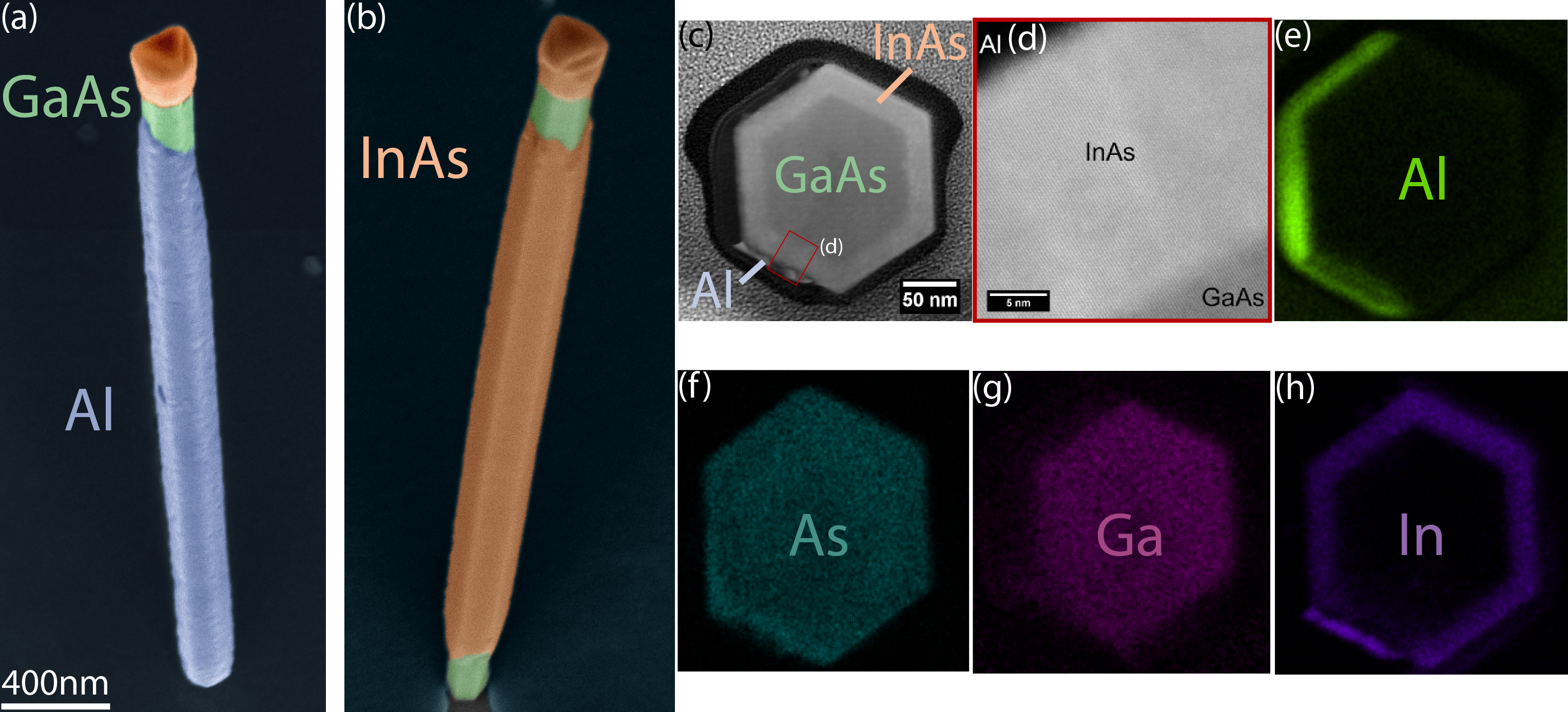}
	\caption[]{False-colored SEM micrographs of the (a) front- and (b) backside of a GaAs/InAs/Al core/shell/half-shell nanowire. (c), (d) ADF-STEM of a cross section of one of the wires from (a), showing the hexagonal crystal structure and a high InAs/Al interface quality. (e)-(h) EDX analysis of the same nanowire.}
	\label{fig:SEM-TEM}
\end{figure*}
Recently, several proposals have emerged that aim to overcome known challenges of trivial-topological phase transitions in hybrid nanowire Josephson junctions, e.g. the need for strong in-plane magnetic fields to open up a helical gap or large trivial charge carrier background contributions. In particular, the so-called Little-Parks effect in nanowire structures that are fully surrounded by a superconducting shell has gained considerable interest due to the much lower requirements in terms of magnetic field strength or the desired length of the Josephson junction channel \cite{Lutchyn2018,Stanescu2018,Vaitieke2020,Lesser2022}. The Little-Parks effect describes the periodic suppression and reemergence of superconductivity due to the acceleration and overheating of the Cooper pair condensate ($T_c\rightarrow\,$0) in a thin-walled cylinder and a winding of the superconducting order parameter associated with it \cite{LittleParks1962}. Such full-shell nanowire devices can display rich physics with analogs to Abrisokov vortices in type-II superconductors \cite{SanJose2023} and have application for flux control in Josephson devices \cite{Svetogorov2023}.

Tunnel- and microwave spectroscopy studies in InAs/Al nanowire/full-shell devices confirmed the existence of the characteristic lobe structure of the Little-Parks effect, but did not find any clear evidence of Majorana zero modes or other topological excitations \cite{Penaranda2020,Valentini2021,Kringhoj2021}. One possible cause for this is an insufficient radial quantization, which not only gives rise to a large number of possible angular momentum states, but might also still allow transport through the bulk of the nanowire \cite{Woods2019}.

\begin{figure*}[!t]
	\centering
\includegraphics[width=0.99\linewidth]{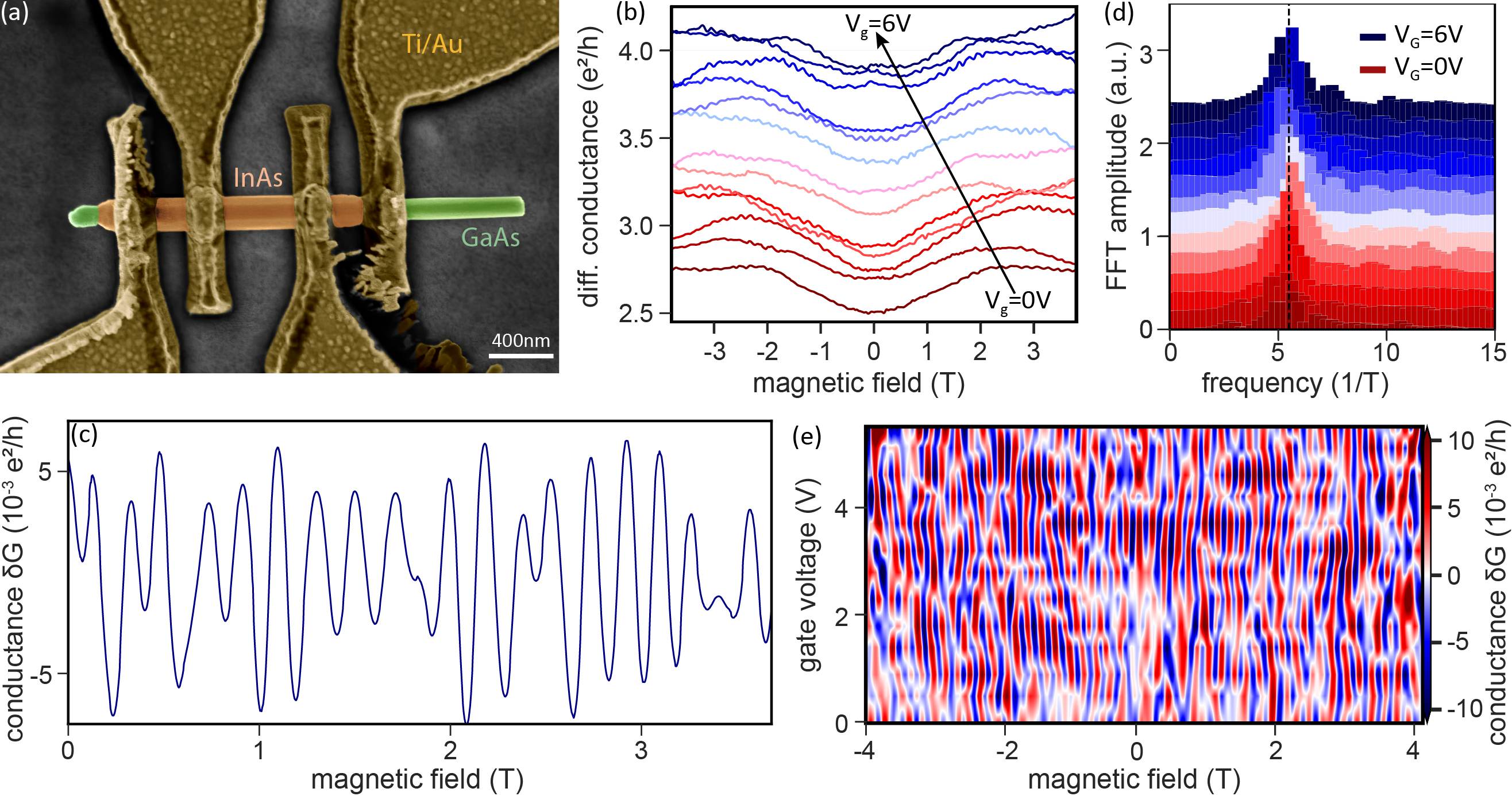}
	\caption[]{(a) Scanning electron micrograph of GaAs/InAs core/shell nanowire (green/orange) equipped with normal conducting Ti/Au contacts (device A). (b) 
	Magnetoconductance in units of $e^{2}/h$ of device A at various gate voltages between $0$ and $6\,\mathrm{V}$ ($0.5\,\mathrm{V}$ steps). (c) Conductance in units of $e^{2}/h$ at $1.7\,\mathrm{K}$ and $V_\mathrm{g}=6\,\mathrm{V}$ after subtracting the slowly varying background signal. (d) Fast-Fourier transformation of the measurements shown in (b). (e) Conductance as a function of magnetic field and gate voltage, showing the evolution of the oscillation phase. }
	\label{fig:Conductance-GaAs-InAs-bare}
\end{figure*}

By combining a wide bandgap core, e.g. GaAs, with a shell made out of a narrow bandgap material such as InAs, it is possible to confine the charge carriers and thereby enhance the radial quantization of the system. These core/shell systems cannot only act as Aharonov-Bohm (AB) type systems \cite{Guel2014,Haas2016,Haas2017,Haas2017a}, but have also been studied in terms of an enlargement of the topological gap due to the steeper band bending, confinement, and wave function engineering and an enhanced spin-orbit interaction \cite{Woods2019, Kloeffel2018,Wojcik2019,Furthmeier2016}. In addition, these systems with an effectively hollow core can be directly analytically mapped to the seminal models of Oreg \textit{et al.} \cite{Oreg2010} and Lutchyn \textit{et al.} \cite{Lutchyn2010}. Here, we take a first step towards these more complex systems by investigating epitaxially-grown GaAs/InAs/Al core/shell/half-shell nanowires that, in contrast to earlier works \cite{Haas2018}, have been fabricated by making use of state-of-the-art techniques such as self-catalyzed selective-area growth and an in-situ and low-temperature deposition of the superconducting shell \cite{Krogstrup2015,Guesken2017,Perla2021}.
First, information on orbital states present in the InAs shell is gained by measuring Aharanov-Bohm type oscillations on bare GaAs/InAs core/shell nanowire. Subsequently, the characteristics of corresponding Josephson junctions are investigated. In particular, the periodic features present in the critical current are studied at various gate voltages, temperatures and magnetic field orientations.

\section {Device details} 

The GaAs/InAs core/shell nanowires are grown selectively by molecular beam epitaxy (MBE) on pre-patterned substrates using the self-catalyzed vapor-liquid-solid method. Details about the nanowire growth can be found in the Methods section. In short, the polymorphic GaAs nanowires are grown selectively, resulting in a typical length of $4\,\mathrm{\upmu m}$ and a diameter of between $100\,\mathrm{nm}$ and $200\,\mathrm{nm}$ (See Supplementary Material for details). Subsequently, the InAs shell is grown by vapor-solid overgrowth of the GaAs core. The total thickness of the InAs shell is about $20\,-\,30\,\mathrm{nm}$. Two growth runs are performed with identical parameters. The nanowires of the first growth run are equipped with Ti/Au normal contacts for magnetotransport measurements (device A). In the second growth run, a $20-30\,\mathrm{nm}$ thick Al half-shell is in-situ deposited on the GaAs/InAs core/shell nanowire. In Figs.~\ref{fig:SEM-TEM} (a) and (b) scanning electron microscope (SEM) images of front- and backside of the Al-half-shell covered GaAs/InAs nanowires are depicted. From this growth run two different Josephson junctions, i.e. devices B and C, have been fabricated and subsequently studied. 

\begin{figure}[t!]
	\centering
\includegraphics[width=0.95\linewidth]{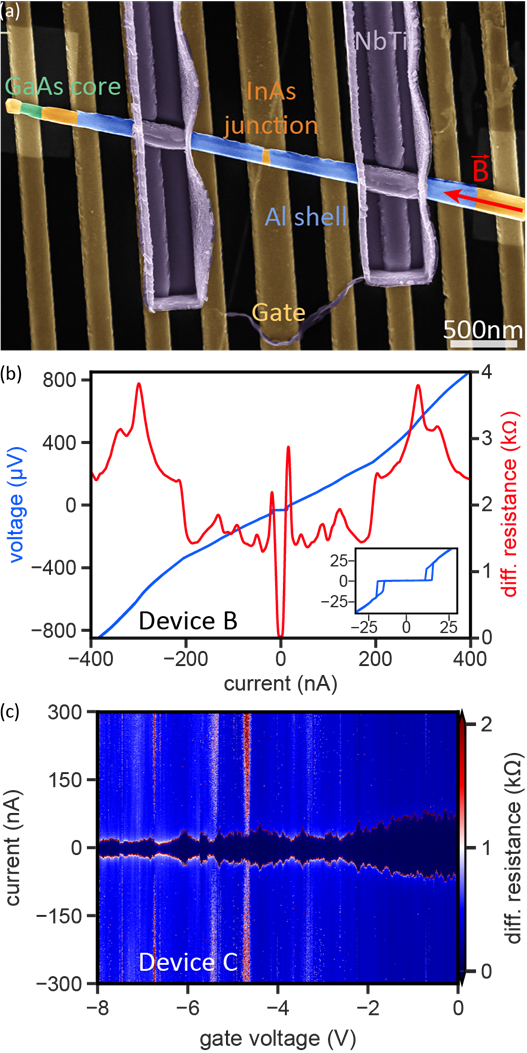}
	\caption[]{(a) False-colored SEM micrograph of an exemplary GaAs/InAs/Al core/shell/halfshell nanowire Josephson junctions with an approx. $140\,\mathrm{nm}$ long channel. (b) Current-voltage characteristics (blue) and differential resistance (red) as a function of bias current of device B at $T=14\,\mathrm{mK}$ and $V_{\mathrm{g}}=0\,\mathrm{V}$. The inset shows detail around the supercurrent branch of the $I-V$ trace, revealing an underdamped behavior. (c) Differential resistance as a function gate voltage and bias current for device C at $T=14\,\mathrm{mK}$.}
	\label{fig:coreshell-junction-overview}
\end{figure}

The structural properties of the GaAs/InAs/Al core/shell/half-shell nanowires are examined using scanning transmission electron microscopy (STEM) to generate both side and cross--section views. Figure~\ref{fig:SEM-TEM} (c) shows an annular dark field scanning transmission electron microscope (ADF-STEM) images of a GaAs/InAs/Al junction cross--section and the corresponding energy dispersive X-ray (EDX) elemental map, which are given in Fig.~\ref{fig:SEM-TEM} (e-h). The thickness of the different components was measured using several cross--sections. Based on that, we determined values of $150\,\mathrm{nm}$, $25\,\mathrm{nm}$, and $20\,\mathrm{nm}$ for the GaAs core, InAs, and Al half shell, respectively. Regarding the semiconductor/superconductor interface shown in Fig.~\ref{fig:SEM-TEM} (d), we observe a high level of quality, similar to what is reported for conventional InAs nanowires with in-situ Al \cite{Krogstrup2015,Perla2021}. Lastly, the EDX maps of Figs.~\ref{fig:SEM-TEM} (e)-(h) depict the chemical composition of the nanowires, confirming no signs of obvious (inter-)diffusion of elements. Further TEM images of the Al/InAs interface and an intensity profile across the GaAs/InAs/Al heterostructure can be found in the Supplemental Materials Section \textbf{I}, Figs.~\textbf{S1} and \textbf{S2}.    

\section{Results and Discussion}

\subsection{Magnetoconductance in GaAs/InAs core/shell nanowires \label{section:normal}}

First, the transport properties of bare core/shell GaAs/InAs NWs are studied. Figure~\ref{fig:Conductance-GaAs-InAs-bare} (a) shows an SEM image of the normal contacted sample (device A) made with wires from the first growth run. In this device, the GaAs core has a diameter of $150\,\mathrm{nm}$ and an InAs shell thickness of $25\,\mathrm{nm}$. To confirm the phase--coherent nature of electronic transport at low temperatures, the conductance of device A was measured under an axially applied magnetic field $B$ at gate voltages ($V_{\text{g}}$) between $0\,\mathrm{V}$ and $6\,\mathrm{V}$. In the obtained curves we observe regular $h/e$--periodic oscillations in the conductance $G$ (cf. Fig.~\ref{fig:Conductance-GaAs-InAs-bare} (b)), which are attributed to AB--type oscillations \cite{Guel2014}. The oscillation amplitude is about $0.5\,\mathrm{\%}$ of the total conductance and does not exhibit a significant increase for higher gate voltages and enhanced electron accumulation \cite{Guel2014}. In order to better resolve the oscillation pattern, the slowly varying background $G_{\mathrm{0}}(B)$ was subtracted from the conductance by means of a moving average fit, resulting in the oscillation signal $\delta G\,=\,G\,-\,G_{\mathrm{0}}(B)$ plotted exemplary in Fig.~\ref{fig:Conductance-GaAs-InAs-bare} (c) for $V_{\text{g}}\,=\,6\,\mathrm{V}$.  The beating pattern in the signal is attributed to small radius variations along the nanowire axis and hence enclosed area $A$, resulting in differences of the enclosed magnetic flux $\Phi\,=\,AB$. In addition, we observe a decrease of the  oscillation amplitude with temperature increase (See Supplemental Material Section \textbf{III}, Figure \textbf{S3}) \cite{Haas2017a} which is consistent with a reduction of the phase-coherence length due to thermally induced scattering events. Figure~\ref{fig:Conductance-GaAs-InAs-bare} (d) shows the fast-Fourier transform (FFT) amplitudes for various gate voltages corresponding to the recorded magnetoconductance oscillations shown in (b). The central frequency peak occurring at $5.5\,\mathrm{T^{-1}}$ implies oscillation period of $\Delta B\,\approx\,182\,\mathrm{mT}$. Assuming a hexagonal cross sectional area of the nanowire and taking into account an estimated oscillation period $\Delta B$, relation $h/e\,=\,\Delta Br^{2}(3\sqrt{3}/2)$ yields a phase--coherent trajectory radius $r\,\approx\,93\,\mathrm{nm}$. Obtained results imply a confined trajectory within the boundaries of InAs shell and shows good agreement with the expected $h/e$-related frequency, hence confirming the origin of observed Aharonov-Bohm-type oscillations in tubular conductor \cite{Guel2014,Haas2017a}. Indeed the radius calculated from the oscillation period agrees very well with the radius of $89\,\mathrm{nm}$ corresponding to a wave function located in the center of the InAs shell. Lastly, we studied the phase rigidity of the oscillations with respect to changes in the applied gate voltage. As can be seen in Fig.~\ref{fig:Conductance-GaAs-InAs-bare} (e), the oscillation phase is shifted when the gate voltage is varied. This behavior is a characteristic feature of AB type oscillations in mesoscopic samples with disorder and can be attributed to different scattering paths around the circumference of the ring if the Fermi wavevector is changed \cite{vanderWiel2003,Haas2016,Guel2014}.

\begin{figure*}[!t]
	\centering
\includegraphics[width=0.99\linewidth]{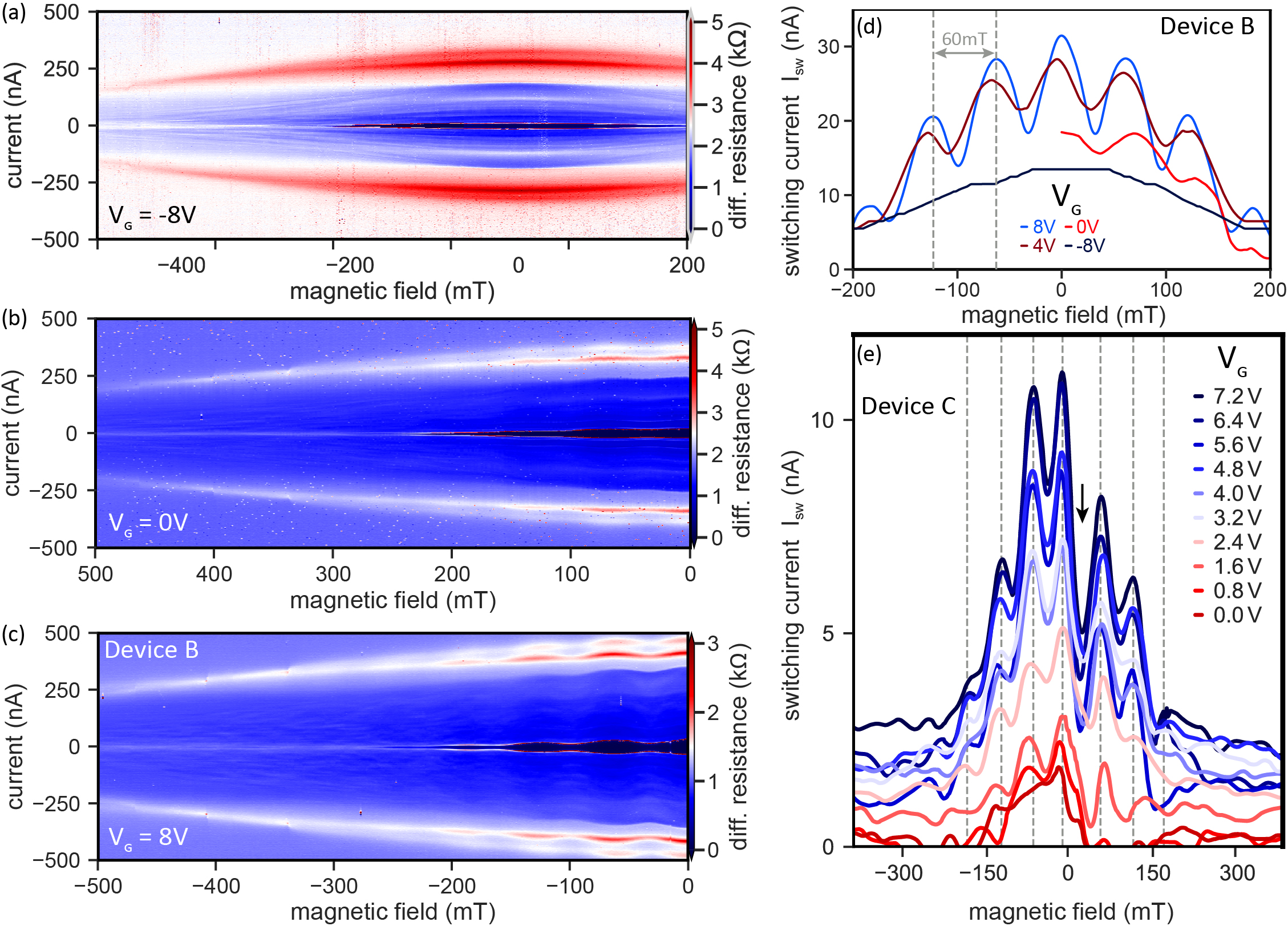}
	\caption[]{(a)-(c) Differential resistance of device B as a function of magnetic field and bias current for $V_{\mathrm{g}}=-8\,\mathrm{V}$, $0\,\mathrm{V}$, and $8\,\mathrm{V}$, respectively. (d) Switching current ($I_{\mathrm{sw}}$) as a function of magnetic field evaluated for the gate voltages shown in (a)-(c). (e) Switching current of device C as a function of magnetic field (swept from negative to positive) at various gate voltages. Device C exhibits a narrow suppressed switching current feature that is hysteretic with respect to magnetic field and likely caused by trapping of vortices in the junction (indicated by an arrow).
    }
	\label{fig:MainTextFig3}
\end{figure*}

\subsection{Josephson junctions based on GaAs/InAs/Al core/shell/half-shell nanowires}

After confirming the presence of phase-coherent transport via tubular states in normal contacted GaAs/InAs core/shell nanowires, we now focus on Josephson junctions based on corresponding core/shell nanowires with in-situ deposited Al half-shells. An example device with an etched Josephson junction is shown in Fig. \ref{fig:coreshell-junction-overview} (a) (device B). As shown in Fig.~\ref{fig:coreshell-junction-overview} (b), for junction device B, a pronounced supercurrent with a switching current of $I_{\mathrm{sw}} (V_{\mathrm{g}}=0\,\mathrm{V})=22\,\mathrm{nA}$, and a normal state resistance ($R_{\mathrm{N}}$) of about $2.2\,\mathrm{k\Omega}$ is observed. Based on the hysteresis in the supercurrent branch (see inset), we determine that the junction is underdamped. In addition, the device exhibits several peaks in the differential resistance ( red trace in Fig.~\ref{fig:coreshell-junction-overview} (b)) that can be interpreted as signatures of multiple Andreev reflections \cite{Octavio1983,Flensberg1988}. For device C, the current-voltage shows a non-zero resistance in the supercurrent branch due to the presence of a parasitic series resistance of $273\,\mathrm{\Omega}$, the subtraction of which leads to a clear Josephson junction behavior, including the presence of well quantized Shapiro steps in AC-driven measurements (see Supplemental Material Section \textbf{IV}, Figure \textbf{S4}). As can be seen in Fig.~\ref{fig:coreshell-junction-overview} (c), for device C, the magnitude of the supercurrent as well as $R_{\mathrm{N}}$ can be altered by adjusting the gate voltage between $V_{\mathrm{g}}=- 8\,\mathrm{V}$ and $0\,\mathrm{V}$. However, no complete pinch-off is achieved. We attribute this to the comparatively thick wire geometry, which lowers the effectiveness of the gate. Indeed, prior devices based on InAs/Al half-shell wires showed a full depletion of the conducting channel for negative gate voltages in the very same gate layout (c.f. Ref.~\cite{Zellekens2020}).

Following earlier experiments on core/shell nanowire devices \cite{Guel2014a,Haas2018}, we focus on the modifications to the electric transport if the device is penetrated by an in-plane magnetic field parallel to the nanowire axis. Figure~\ref{fig:MainTextFig3} (a) shows the change in differential resistance of the nanowire Josephson junction device B for an applied axial in-plane magnetic field and $V_{\mathrm{g}}=-8\,\mathrm{V}$. In this configuration, the device exhibits the typical iris-shaped closing of the superconducting gap (see e.g. Ref. \cite{Zellekens2020}) without any sign of additional quantum (interference) effects. However, for $V_{\mathrm{g}}=0\,\mathrm{V}$ (see Fig.~\ref{fig:MainTextFig3} (b)), the response of the junction changes significantly. Here, both the supercurrent and the sub-gap Andreev transport start to show pronounced flux-periodic oscillations on top of the aforementioned closing of the superconducting gap (in the following referred to as background). The oscillations become even more prominent is raised to 
  large positive gate voltage ($V_{\mathrm{g}}=8\,\mathrm{V}$, c.f. Fig.~\ref{fig:MainTextFig3} (c)). Figure~\ref{fig:MainTextFig3} (d) shows the extracted switching current ($I_{\mathrm{sw}}$) of the junction, i.e. the current value above which the device starts to exhibit a finite resistance, as a function of magnetic field at several gate voltages for device B. For all gate voltages, we observe the previously mentioned iris-shaped background that is associated with the closing of the superconducting gap. However, above $V_{\mathrm{g}}=0\,\mathrm{V}$, flux-periodic oscillations of $I_{\mathrm{sw}}$ emerge, with a maximum amplitude at $V_{\mathrm{g}}=8\,\mathrm{V}$ of approx. $1/4$ of the zero-field background current. We note that the results are similar to a recent report in InAsSb/Al junctions where the surface accumulation layer provides a tubular channel \cite{Stampfer2022}.
\begin{figure*}[!t]
\includegraphics[width=0.98\linewidth]{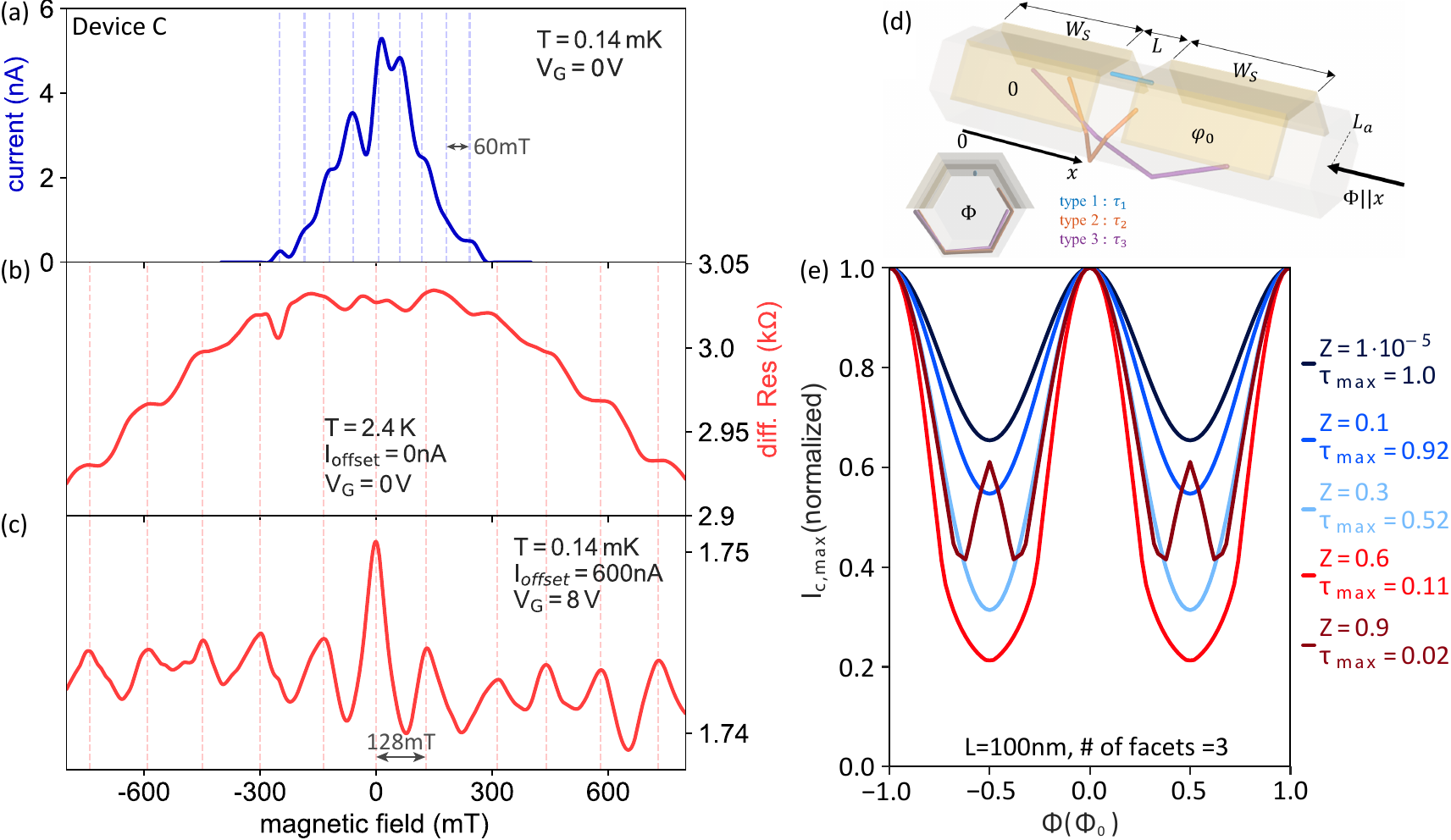}
	\caption[]{(a) Switching current of device C measured at $V_{\mathrm{g}}=8\,\mathrm{V}$ and at $T=14\,\mathrm{mK}$. (b) Differential resistance measured at an elevated temperature of $2.4\,\mathrm{K}$ and the same gate condition as in (a). (c) Differential resistance at $T=14\,\mathrm{mK}$ and $V_{\mathrm{g}}=8\,\mathrm{V}$ with a DC current bias of $600\,\mathrm{nA}$, probing transport outside the bias window of the superconducting gap. (d) Schematic showing of three types of trajectory on the surface of the hexagonal cross section nanowire. The displayed type-2 and type-3 trajectories contribute an AB phase while the type-1 trajectory does not. (e) Example simulation result for a range of dimensionless barrier stengths $Z$, showing the critical current as a function of magnetic flux, with $\Phi_0=h/2e$ the magnetic flux quantum. Each trace is also labelled with the maximum value of transparency ($\tau_{max}$) evaluated for all found trajectories.
    }
	\label{fig:MainTextFig4}
\end{figure*}

In the case of the Josephson junction the highly visible oscillations allow simple extraction of the period, as shown in Fig.~\ref{fig:MainTextFig3} (d), revealing $\Delta B \approx 60\,\mathrm{mT}$ in device B. This value corresponds to an effective radii of $r_{h/e} \approx 158\,\mathrm{nm}$ assuming $h/e$ periodic oscillations or $r_{h/2e} \approx 112\,\mathrm{nm}$ assuming $h/2e$ periodic oscillations, for a hexagonal cross section, as in Section \ref{section:normal}. Based on an average total wire diameter of $300\,\mathrm{nm}$, with an Al thickness of around $t_{\mathrm{Al}}\,= 25\,\mathrm{nm}$ and an InAs thickness of $t_{\mathrm{InAs}}=25\,\mathrm{nm}$ and the fact that the maximum of the probability function is situated in the middle of the InAs shell \cite{Guel2014}, we find a good correspondence to the calculated effective radius $r_{h/2e}$. The relatively large difference to the cross sectional dimensions of device A can be explained by the significant spread of nanowire GaAs core diameters between different arrays with holes of different diameters and pitches on the very same growth substrate (see Section \textbf{II}, Table \textbf{SI} and \textbf{SII} in the Supplemental Material for a more detailed analysis of the different nanowire dimensions across the growth substrates).
\begin{figure*}[!t]
\includegraphics[width=0.98\linewidth]{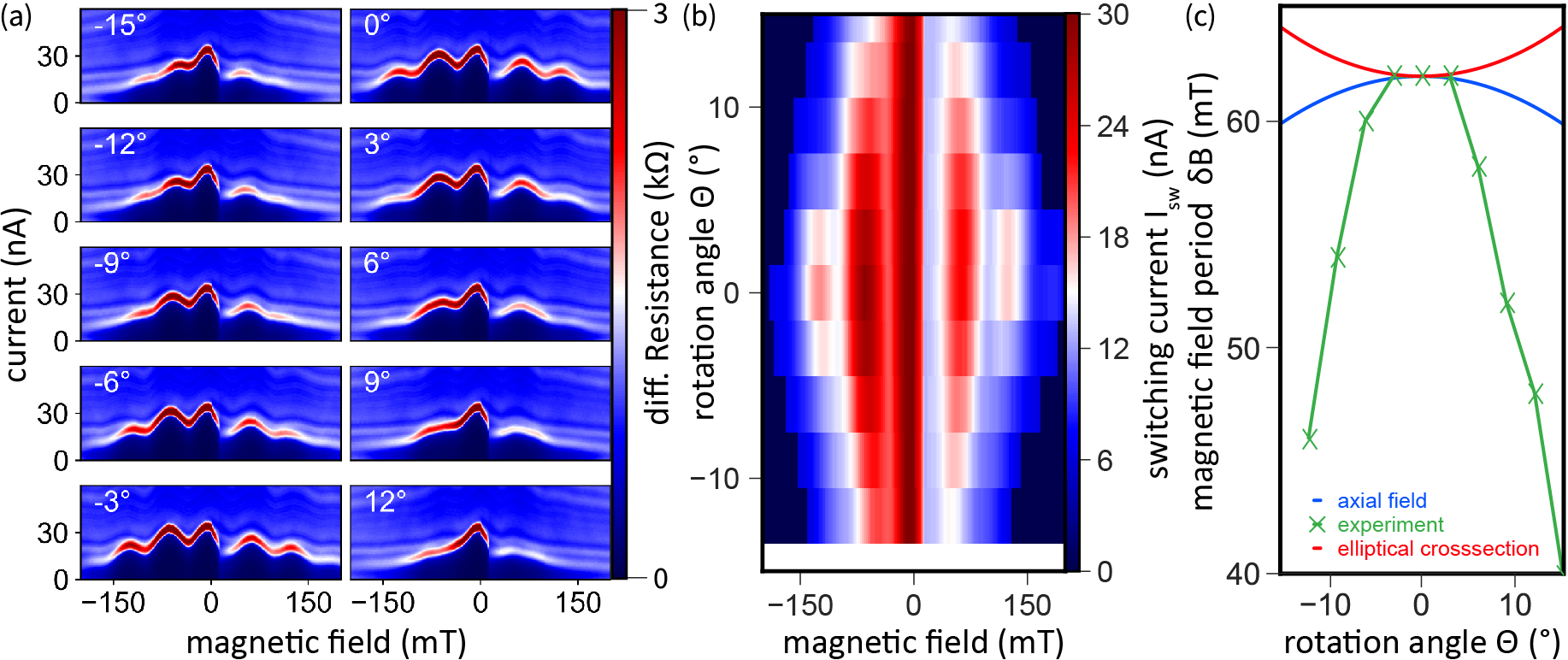}
	\caption{Data of field angle dependence for device B at $V_{g}=4.0\,$V. (a) These plots show the raw data for the plot in (c). The color bar is $dV/dI$ with units of $\mathrm{k\Omega}$. (b) Switching current $I_{sw}$ as a function of field angle $\theta$ from which the magnetic field period $\Delta B$ is extracted and plotted in (c). (c) The decreasing $\Delta B$ is compared with expectations based on flux threading the nanowire axis (red dashed line) and threading the elliptical cross section (solid blue line).}
	\label{fig:Rotation_Data}
\end{figure*}

Figure~\ref{fig:MainTextFig3} (e) provides an overview of the gate- and magnetic field-dependent evolution of the switching current for device C. This device exhibits a region of suppressed $I_{\mathrm{sw}}$ (indicated by the arrow in Fig.~\ref{fig:MainTextFig3} (e)) that is hysteretic in magnetic field and we attribute to trapping of vortices within the junction \cite{YosukePRL2022}. Similar to results of device B, the depletion-induced damping of the flux-periodic oscillations is much stronger than the one of the background supercurrent. Here, comparable radii as to device B are evaluated from the oscillation period of $60\,\mathrm{mT}$ with $r_{h/e} \approx 163\,\mathrm{nm}$  or $r_{h/2e} \approx 115\,\mathrm{nm}$ assuming $h/e$ or $h/2e$ periodic oscillations, respectively. Contrary to the results for the normal conducting samples shown in Fig.~\ref{fig:Conductance-GaAs-InAs-bare} (e),  fixed positions of the minima and maxima (see vertical lines in Fig.~\ref{fig:MainTextFig3} (e)) of the oscillations are observed. This resilience against changes in the applied gate voltage is a typical signature of quantum interference effects protected by time-reversal symmetry, such as $h/2e$-periodic Al'tshuler-Aronov-Spivak oscillations in normal conducting, mesoscopic samples \cite{Altshuler1981}, for which any effect of random scattering loops caused by modifications of the Fermi wave vector $k_{\mathrm{F}}$, and thus the aforementioned additional phase shift, is cancelled \cite{Altshuler1981}.

To gain a more detailed picture of the origin of the observed flux-periodic supercurrent oscillations in connection to the geometrical conditions, we investigate the transport in device C at temperatures above $T_{\mathrm{c}}$ of the aluminum half-shell and at bias voltages outside of the induced proximity gap, as shown in Figs.~\ref{fig:MainTextFig4} (a)-(c), for $V_{\mathrm{g}}=0\,\mathrm{V}$, $V_{\mathrm{g}}=0\,\mathrm{V}$ and $V_{\mathrm{g}}=8\,\mathrm{V}$, respectively. Figure~\ref{fig:MainTextFig4} (a) shows the evaluated switching current for $T=14\,\mathrm{mK}$ for comparison with traces in (b) and (c). Figure~\ref{fig:MainTextFig4} (b) shows the differential resistance ($dV/dI$) measured at $T=2.2\,\mathrm{K}$ and at zero DC current ($I_{\mathrm{offset}}=0$), where the aluminium half-shell forming the junction is in the normal state ($T_{\mathrm{c}}\sim 1\,\mathrm{K}$). Comparing the peak separation in (a) and (b), indicated by the dashed lines, one finds that the period of the oscillations approximately doubles if the transport takes place above the critical temperature of the aluminium half-shell. We also investigate how the system responds at temperatures for which the shell is superconducting ($T \ll T_{\mathrm{c}}$) if subjected to bias currents that drive the transport into the dissipative regime. Using a superposition of a constant DC offset current of $I_{\mathrm{offset}}=600\,\mathrm{nA}$ and a small AC lock-in signal, the resistivity oscillations shown in Fig.~\ref{fig:MainTextFig4} (c) reveal once more a prominent increase in their periodicity, as compared to the low-temperature and zero-bias response in Fig.~\ref{fig:MainTextFig4} (a). We note that the period of oscillations in the superconducting transport is approximately $60\,\mathrm{mT}$ as compared to about $128\,\mathrm{mT}$ in the normal transport. Assuming that superconducting state oscillations are $(h/2e)$ periodic and normal oscillations are $(h/e)$ periodic this corresponds to radii of $115\,\mathrm{nm}$ and $112\,\mathrm{nm}$, respectively. 

The findings for the gate-, temperature- and current-dependent evolution of the junction response indicate a clear correlation between the special device properties, i.e. the complex geometric topology, and the $h/2e$-oscillations. Furthermore, the observation of supercurrent oscillations goes beyond the earlier results reported in Refs.~\cite{Guenel2012,Guel2014a,Haas2018} that were obtained on GaAs/InAs core/shell nanowire-based Josephson junctions with ex-situ contacts and global backgates and which could only observe the flux-periodic oscillations in the normal state resistance $R_{\mathrm{N}}$. In this context, we interpret the results as an interference effect that takes place on the level of flux-periodic Andreev reflection/Andreev bound states within the nanowire. Following this picture, we assume that Andreev transport processes enclosing the threading flux can occur around the perimeter of the nanowire in regions without aluminium shell. This in turn is a strong indication that the InAs shell is fully proximitized in the aluminium covered regions of the device. We expect that two effects may play a role in the device as discussed below. The first is based on the Little-Parks effect in an inhomogenous cylinder, in which most of the modulation of the superconducting order parameter occurs in the area with weaker superconductivity \cite{Nikulov1998}. In this picture, the proximitized InAs shell on the side of the nanowire opposite the aluminium half-shell (i.e. the underside of the nanowire) forms an extended, additional weak-link that can support a non-zero non-dissipative current. Upon application of a field along the wire axis a screening current is induced around the wire circumference with a magnitude fixed by the current-phase relationship of this extended weak-link. The condensate velocity due to the screening current can then cause modulation of the critical temperature of the Al half-shell akin to the Little-Parks effect, which in turn will modify the current that can be supported by the weak-link. In the conventional Little-Parks effect the screening current varies linearly with applied flux in contrast to the proposed sinusoidal variation due to the weak-link formation \cite{SharonSciRep2016}. However, devices fabricated with normal contacts and no etched Josephson junction reveal no significant modulation of the superconducting transport (see Supplemental Material Section \textbf{V}, Figure \textbf{S6}), indicating that the above discussed effect, while possible, is likely not the primary cause of the strong modulation observed in the Josephson junction devices. Furthermore, the observed gate dependence gives a strong indication that the effect is likely specific to transport in vicinity of the defined junction.

In a second interpretation, the effect arises from the Josephson junction transport. Here, the oscillations are based on superimposed Andreev reflection trajectories that loop around the perimeter of the InAs shell and enclose magnetic flux. A similar interpretation has recently been employed to explain the magnetic response of topological insulator nanowire Josephson junction devices in which surface states dominate the transport \cite{HimmlerPRR2023}. To illustrate this effect, we utilize a simplified semi-classical model following the work of Himmler \textit{et al.}~\cite{HimmlerPRR2023} in which the different current contributions are integrated over all possible ballistic trajectories \cite{OstroukhPRB2016}. The model used here closely follows the outline by Himmler \textit{et al.} in Ref.~\cite{HimmlerPRR2023} and is discussed in detail in the methods section with additional plots in the Supplemental Material Section \textbf{VI}. Each trajectory represents an electron and hole path, with Andreev reflections occurring at each superconductor interface. Three different types of trajectories can be identified, which differ by the relationship between the angle of incidence ($\theta_{\mathrm{i}}$) at the two relevant superconductor interfaces and the trajectory angle with respect to the wire growth axis ($\theta_{\mathrm{t}}$). Examples of the three trajectory types are shown in Fig.~\ref{fig:MainTextFig4} (a). Type 1 trajectories pass between the two short edges of the superconducting leads at the junction and are therefore equal to the transport situation in a planar system. Type 2 trajectories pass between a short edge of a contact and a long contact edge that is parallel with the nanowire growth direction. Finally, type 3 trajectories pass between two long contact edges. In all cases, it is possible to find trajectories that wind around the nanowire perimeter, enclose a magnetic flux and induce an AB-type phase shift. We utilize expressions for the Andreev reflection probability taken from the work of Mortensen \textit{et al.} in Ref.~\cite{MortensenPRB1999} to account for the relation between incident angle ($\theta_{\mathrm{i}}$) and the transparency of the channel $\tau$. In Fig.~\ref{fig:MainTextFig4} (e), we use the Blonder and Tinkham model~\cite{BlonderTinkhamPhysRevB1983} to simulate the device response for different tunnel barrier strength $Z$. For typical values for nanowire junctions with an epitaxial superconducting shell \cite{Krogstrup2015,Zellekens2020a}, we obtain clear $h/2e$ oscillations that are in accordance with our experimental findings. In the case of very high transparencies, additional non-topological $h/4e$ oscillations can be observed (see Supplementary Material Section \textbf{VI}), as has been reported in topological insulator nanowires \cite{HimmlerPRR2023}. However, we have not detected any signs of such features in our devices. 

Lastly, we investigate the relation between the supercurrent oscillations and the orientation of the magnetic field, i.e. the effect of an off-axis misalignment of the applied magnetic field, on the flux periodicity in device B (see Fig.~\ref{fig:Rotation_Data}). The oscillation period ($\Delta B$) is extracted from the separation of two peaks close to zero field at $V_{\mathrm{g}}=4.0\,\mathrm{V}$ and for fields applied in-plane and at an angle $\theta$ to the nanowire axis (see Fig.~\ref{fig:Rotation_Data} (a) and (b) for the experimental data). As shown in Fig.~\ref{fig:Rotation_Data} (c), we observe that an off-axis orientation of the magnetic field leads to an anomalous decrease of the oscillation period. For comparison, the dashed red curve in Fig.~\ref{fig:Rotation_Data} (c) corresponds to the expected oscillation period assuming that only the axial component of the magnetic field produces the oscillations. This prediction follows a cosine dependence that is typical for conventional AB-type oscillations, universal conductance fluctuations and weak localization \cite{Haas2016,Jespersen2015,Dong2010}. We also consider the expected periodicity if the oscillation period was fixed via the flux penetrating the elliptical cross section of the nanowire (shown as a solid blue line in Fig.~\ref{fig:Rotation_Data} (c)), which similarly does not reflect the observed dependence. The anomalous angular dependence of the oscillations is a clear indication that the given system can not be described as a simple superconducting AB-type system. However, a similar atypical modification of the oscillation period is observed in the Little-Parks effect in InAs nanowires with full-shells of epitaxial aluminium (see Supplemental Material Section \textbf{VII}) and can most likely be explained by the additional depairing effect of the out-of-plane magnetic field component (see \cite{Vekris2021}). Here, the superconducting gap, which acts as the envelope for all Josephson junction related transport features, is rapidly suppressed. This creates an upper cut-off condition for the supercurrent oscillations and thus an apparent shift of the oscillation maxima towards lower fields.

\section{Conclusions}

The study of topological quasiparticles in nanowire-superconductor hybrid structures remains a challenging task due to the strict requirements on the device properties. However, more advanced device layouts such as full-shell nanowire Josephson junctions are a promising approach to loosen the tight margins and making the topological phase space more easily accessible. Here, an essential prerequisite is a fine control of the position of the electron wave function within the conductive area of the nanowire. This can be achieved by employing GaAs/InAs core/shell nanowires, where the electrons are confined in the narrow InAs quantum well.
Indeed, our GaAs/InAs core/shell nanowires equipped with normal contacts revealed pronounced $h/e$ flux-periodic conductance oscillations which confirm the presence of a tubular channel in the InAs shell. Moreover, in nanowires contacted with a pair of closely spaced superconducting Al electrodes, $h/2e$ periodic supercurrent oscillations are observed. By subjecting the system to elevated temperatures and high currents, thereby surpassing the induced proximity gap, we can dynamically switch the oscillation period from $h/2e$ to $h/e$, transitioning from a two-particle to a single-particle effect. This, paired with the possibility to suppress the oscillatory behavior of the supercurrent by means of an asymmetrically applied gate voltage, as well as the fact that no such oscillations are observed in devices without an etched junction channel, is a clear indication that the effect originates from the superposition and interference of different Andreev reflection processes within the nanowire Josephson junction and that the complex geometric topology can indeed be imprinted onto the transport in the device.  The latter is further confirmed by a semi-classical simulation that investigates the relation between interface transparency, contact geometry and the resulting oscillations. Lastly, we provide compelling evidence for an anomalous relation between magnetic field orientation and oscillation period that can be explained by an enhanced depairing effect and the rapid suppression of the superconducting gap.
Our measurements show that GaAs/InAs core/shell nanowires are a promising platform for more advanced device concepts based on the Little-Parks effect and to study topological quasiparticles in nanowire-superconductor hybrid structures. In addition, they provide a template for further investigations regarding the relation between device geometry and the supercurrent oscillations, e.g. the effect of very long junction channels or very short or asymmetric Al contact regions.

\section*{Methods}

\subsection*{Epitaxial growth}

The GaAs/InAs core/shell nanowires are grown by molecular beam epitaxy on pre-patterned (111)-Si substrates using the self-catalyzed vapor-liquid-solid method. Prior to the growth, the (111) Si substrates are covered with a $20\,\mathrm{nm}$ thick thermally grown $\mathrm{SiO_{2}}$ layer. This is subsequently patterned via electron beam lithography and reactive ion etching (RIE) to generate a set of hole arrays with different diameters ranging from $d = 40\,\mathrm{nm}$ to $80\,\mathrm{nm}$ and varying hole pitches of $p = 0.5, 1, 2$ and $4\,\mathrm{\upmu m}$. The preparation procedure can be divided into seven process steps (see Ref. \cite{Jansen2020} for a more detailed description). In step (1) the (111) Si substrates are covered by a $20$\,nm thick thermally-grown SiO$_2$ layer as a mask for selective area growth. Subsequently, an approximately $220\,\mathrm{nm}$ thick PMMA-950\,K resist layer is spin-coated (2) and afterwards in step (3), the layout is transferred to the PMMA by electron beam lithography. Then, the PMMA layer is developed with AR-600-55 for $70\,\mathrm{s}$ and subsequently the process is stopped by sample immersion in isopropanol for $3$\,min. The first etching process is conducted by RIE to thin the $\mathrm{SiO_{2}}$ in the holes down to $1-2\,\mathrm{nm}$ thickness using a $\mathrm{CHF_{3}}$ gas flow of $50\,\mathrm{sccm}$ and a RF bias power of $200\,\mathrm{W}$ (4). After that, the PMMA layer is removed using acetone, and the patterned substrates are chemically cleaned using Piranha solution and oxygen plasma for $10\,\mathrm{min}$ each (5). Directly before loading the sample into the MBE system, it is chemically wet-etched using diluted ($1\mathrm{\%}$) HF for another $60\,\mathrm{s}$ (6) in order to open the remaining $1-2\,\mathrm{nm}$ of $\mathrm{SiO_{2}}$ in the holes of the pre-patterned substrate. Immediately after the HF dip the samples are loaded in the III/V MBE chamber (7). 

The growth procedure of the GaAs/InAs core/shell NWs is initiated by exposure of the pre-patterned substrate to a Ga-flux (about $0.11\,\mathrm{\mu m/h}$) for $10\,\mathrm{min}$ at a growth temperature of $620\,\mathrm{^\circ C}$. During this step, Ga droplets are formed in the holes and act as catalyst particles. Subsequently, the As shutter (BEP(As) = $5\times 10^{-6} \,\mathrm{mbar}$) is opened to start the growth of the GaAs NW core for a duration of about $105\,\mathrm{min}$, which leads to an average NW length of about $4\,\mathrm{\mu m}$. Since the Ga flux is kept constant all the time, the GaAs core growth is polymorph, i.e. containing segments of wurtzite and cubic crystal stucture along the NW axis. More details on the growth dynamics of such polymorph GaAs NWs on prepatterened substrates are presented in Ref.~\cite{Jansen2020}. At the end of the GaAs core growth, the sample is exposed for an additional $20\,\mathrm{min}$ to the As flux in order to consume the catalyst droplet on top of each nanowire. The GaAs nanowires have a in a typical length of $4\,\mathrm{\mu m}$ and a radius of $100\,\mathrm{nm}$ to $200\,\mathrm{nm}$. Here, the radius is defined as the distance from the center of the GaAs core to one of the corners of the hexagonal InAs shell.

Subsequently, the InAs shell growth is performed by vapour-solid overgrowth of the GaAs core. Here, an In flux of about $0.1\,\mathrm{\mu m/h}$ and the same As flux as for the GaAs core growth is applied. The growth temperature is set to $450\,\mathrm{^\circ C}$. In contrast to older studies (see \cite{Rieger2012}), we deposit the InAs using a sequence of four alternating cycles, each composed of $5\,\mathrm{min}$ InAs growth (e.g. both shutters, In and As open) and $5\,\mathrm{min}$ As-stabilized growth break (e.g. only As open) in order to avoid the formation of a parasitic InAs crystallite on top of the nanowires. The resulting total thickness of the InAs shell is about $20-35\,\mathrm{nm}$.

At the end of the InAs shell growth the sample is cooled-down and in-situ transferred to the metal deposition chamber to deposit the Al half shell. No sample rotation is applied for this step and the growth temperature is about $-115\,\mathrm{^\circ C}$ in order to achieve a smooth Al coverage of the InAs surface. The overall half-shell thickness is about $20-30\,\mathrm{nm}$. Scanning electron micrographs of the grown core/shell nanowires covered with an Al half-shell are shown in Figs.~\ref{fig:SEM-TEM}(a) and (b). The thickness of InAs shell can be determined directly from these images due to incomplete coverage of the GaAs core at the bottom or top part of the nanowire.

\subsection*{Device fabrication}
After the growth, nanowires were transferred from the growth substrate onto a Si(100) substrate. Here, a SEM-based micromanipulator, equipped with a tungsten tip, is used in order to achieve a high level of accuracy. Depending on whether bare core/shell GaAs/InAs or GaAs/InAs/Al half-shell nanowires were characterized, different types of substrates were utilized. In the case of bare core/shell GaAs/InAs nanowire, highly n-doped Si(100) substrates covered with $150\,\mathrm{nm}$ of $\mathrm{SiO_{2}}$ were used. This layer is employed as the gate dielectric, through which the electron density is varied in the nanowire by applying a global back-gate voltage $V_{\mathrm{g}}$.  By means of electron beam lithography and subsequent metallization, Ti/Au contact fingers were defined. Before deposition of the contact fingers Ar$^\mathrm{+}$ cleaning was applied. 

To study the dynamics of Josephson junctions in GaAs/InAs/Al half-shell nanowires, we use the measurement platform introduced in \cite{Zellekens2020a, Perla2021}. The latter is based on highly-resistive Si substrates with pre-defined surface gates. The Josephson junction itself is connected through a coplanar-waveguide-type transmission line to an on-chip bias tee, made of an inter-digital capacitor and a planar coil. All parts, including the surrounding ground plane, are made of $80\,\mathrm{nm}$ thick TiN. To ensure an ohmic connection between NbTi contacts and the Al shell a fast atom gun (Matsusada Precision FAB-110) was used prior to metal deposition in order to remove the aluminium oxide. The contact separation is chosen to be at least $1.5\,\mathrm{\mu m}$ in order to reduce the impact of the wide-gap superconductor NbTi on the transport through the system. 

\subsection*{Transmission electron microscopy}

For the side view analysis, the nanowires were transferred from growth arrays to carbon grids by gently rubbing the two surfaces. The cross sections were prepared using focused ion beam (FIB) on nanowires transferred to Si substrates by the very same method. The subsequent TEM analysis was carried out using doubly corrected JEOL ARM 200F and JEOL 2100 microscopes, both operating at $200\,\mathrm{kV}$. The EDX measurements were carried out using an Oxford Instruments $100\,\mathrm{mm}^{2}$ windowless detector installed within the JEOL ARM 200F.

\subsection*{Electrical measurements}

Low-temperature magnetotransport measurements of the core/shell nanowires with normal contacts were carried out in a variable temperature insert cryostat at temperatures between $T\,=\,1.7\,\mathrm{K}$ and $20\,\mathrm{K}$, under an axially applied magnetic field up to $B\,=\,4\,\mathrm{T}$. Biasing with a  current $I\,=\,20\,\mathrm{nA}$ between the two outer contacts, measurements were performed in four-terminal configuration by recording the voltage drop with standard lock-in technique between two inner contacts with measurement segment length of $1\,\mathrm{\upmu m}$. 

Measurements on Josephson junction devices were performed using a four-terminal setup to the NbTi leads contacting the nanowire. Measurements were performed in a dilution refridgerator with a base temperature of $14\,\mathrm{mK}$, equipped with a two axis superconducting vector magnet allowing accurate alignment of field parallel to the wire growth axis. Electrical transport measurements were performed using current bias with custom battery powered electronics (TU Delft QT—IVVI rack) (https://qtwork.tudelft.nl/~schouten/ivvi/index-ivvi.htm). Measurements of $dV/dI$ at elevated temperature and higher currents in Figs. \ref{fig:MainTextFig4} (b) and (c) were performed using standard lock-in techniques with a signal recovery 7270 DSP lock-in amplifier and again the battery powered IVVI electronics. In contrast to prior results on both normal and superconducting samples \cite{Guenel2012,Guel2014a,Haas2018}, the thin layer of epitaxial Al was rather sensitive to out-of-plane components of the applied field. To address this issue, we performed calibration procedures for both junction devices in which we rotated the magnetic field around the nanowire and thus locate the orientation for which the critical field is at its maximum. 

\subsection*{Semiclassical model for Josephson junction magnetic response}

Here, we employ a semi-classical model adapted from the work of Himmler \textit{et al.}~\cite{HimmlerPRR2023}, itself adapting a model from Ostroukh \textit{et al.} \cite{OstroukhPRB2016}. The model considers classical self-retracing trajectories ($\Gamma$) due to pure retro-reflections at interfaces of the superconducting shell in vicinity of the junction. No trajectories arising from normal reflections are considered and as such the model is a significant over-simplification, but nonetheless an useful tool to illustrate the underlying physics of the observed magnetic response. For a full discussion of the model we direct readers to the source in Ref.~\cite{HimmlerPRR2023}, which is only briefly summarized here. 

We consider the system pictured in Fig.~\ref{fig:MainTextFig4} (d). Transport is considered to occur on the surface of the hexagonal cross section nanowire with three of its six facets covered with superconducting aluminium. The hexagonal cross section of the wire has an apothem of $L_{\mathrm{a}}=100\,\mathrm{nm}$. We consider a junction of length $L=100\,\mathrm{nm}$ and sections of superconducting shell leads of length $W_{\mathrm{S}}=1000\,\mathrm{nm}$. We find all trajectories that pass between the two superconducting regions and assume that each trajectory contributes a current $j(\Gamma)$. The total current is found by integration over all contributions $j(\Gamma)$ for all paths. The junction is assumed to be in the short junction limit, $L\ll \xi = hv_{\mathrm{F}}/\Delta_{\mathrm{0}}$ ,with $v_\mathrm{F}$ the Fermi velocity and $\Delta_0$ the superconducting gap energy. It is useful to define the edge length $L_{\mathrm{edge}}=2L_{\mathrm{a}}/\sqrt{3}$, the perimeter of the nanowire $P=6L_{\mathrm{edge}}$ and the perimeter distance covered by superconductor $C=3L_{\mathrm{edge}}$.

In practice, following Himmler \textit{et al.}~\cite{HimmlerPRR2023}, we take a cut through the center of the normal junction region (at $x=W_{\mathrm{S}}+(L/2)$) and characterise all trajectories with a coordinate $s$ along the cut and the axial wave number $k_{\mathrm{s}}$. The total current is then given by 

\begin{align*}
    I&=\frac{1}{2\pi}\int ds\int dk_{\mathrm{s}} j(s,k_{\mathrm{s}}),\\
    I&=\frac{k_{\mathrm{F}}}{2\pi}\int ds\int d\theta_{\mathrm{t}} \cos(\theta_{\mathrm{t}}) j(s,\theta_{\mathrm{t}}) \; ,
\end{align*}

\noindent where $\theta_{\mathrm{t}}$ is the angle of the trajectory with respect to the cut line in the $z$ direction around the nanowire cross section perimeter. Each path carries a current given by the expression for the zero-temperature current phase relationship of a short channel junction \cite{BeenakkerPhysRevLett1991,Kalpwijk2004},
\begin{equation*}
    j(\theta_{\mathrm{t}})=\frac{e\Delta}{4\hbar}\frac{\tau\sin(\varphi_{0}-2\gamma)}{\sqrt{1-\tau\sin^{2}(\varphi_{\mathrm{0}}/2 - \gamma)}},
\end{equation*}
where $\tau$ is the transparency for the given trajectory, and $\varphi_{\mathrm{0}}-2\gamma$ is the gauge-invariant phase difference \cite{tinkham2004Book} which includes the Aharonov-Bohm phase term,
\begin{equation*}
    \gamma=\frac{e}{\hbar}\int_{\Gamma}ds\cdot A=n\pi\frac{\Phi}{\Phi_{\mathrm{0}}},
\end{equation*}
with $\Phi$ the magnetic flux and $\Phi_0=h/2e$ the magnetic flux quantum.
In practice, the term $\gamma$ is zero except for trajectories which cross between the source and drain superconductor while traversing the normal region on the underside of the nanowire. Much like Himmler \textit{et al.}~\cite{HimmlerPRR2023}, we identify three distinct types of trajectory that differ in details of the angles of incidence at the superconductor interfaces (see examples in Fig.~\ref{fig:MainTextFig4} (d)), these are:

\begin{itemize}
\item Type 1 - Trajectories that begin at $s=W_{\mathrm{S}}$ and end at $s=W_{\mathrm{S}}+L$. These trajectories have the same angle of incidence $\theta_{\mathrm{i}}$ on both superconducting leads.
\item Type 2 - Trajectories with one end at the short edge of the superconducting leads (at $s=W_{\mathrm{S}}$ or $s=W_{\mathrm{S}}+L$) and one end along the long edge of the superconducting lead ($z=0$ or $z=C$).
\item Type 3 - Trajectories that both begin and end at the long edge of the superconducting leads ($z=0$ or $z=C$).
\end{itemize}

\noindent These three types differ in terms of the angle of incidence $\theta_{\mathrm{i}}$ of the trajectory with the superconductor interface and its relationship with the trajectory angle $\theta_{\mathrm{t}}$. In contrast to the work of Himmler \textit{et al.} \cite{HimmlerPRR2023}, we use expressions for the angular dependence of Andreev reflection probabilities at semiconductor-superconductor interfaces taken from Mortensen \textit{et al.} \cite{MortensenPRB1999} in order to evaluate the transparencies ($\tau$) of each trajectory. From Ref.~\cite{MortensenPRB1999}, we describe the transparency as $|A(\theta_{\mathrm{i},\mathrm{source}})A(\theta_{\mathrm{i},\mathrm{drain}})|$ where $A(\theta_{\mathrm{i},\mathrm{m}})$ is the Andreev reflection probability at the start or end interface of the trajectory given as
\begin{align*}
    A(\theta_{\mathrm{i},\mathrm{m}})&=\frac{1}{\tilde{E}^{2}+(1-\tilde{E}^{2})(1+2Z_{\mathrm{eff}}^{2}(\theta_{\mathrm{i},\mathrm{m}}))^{2}}\:\mathrm{for}\: \theta_{\mathrm{i},\mathrm{m}}<\theta_{\mathrm{c}} \, , \\
    A(\theta_{\mathrm{i},\mathrm{m}})&=0\; \:\mathrm{for}\: \; \theta_{\mathrm{i},\mathrm{m}}\ge\theta_{\mathrm{c}},
\end{align*}
where $\tilde{E}=E/\Delta_{\mathrm{0}}$ is the normalized energy and $Z_{\mathrm{eff}}$ is an effective tunnel barrier strength given as
\begin{equation*}
    Z_{\mathrm{eff}}(\theta_{i,m})=\sqrt{\kappa(\theta_{\mathrm{i},\mathrm{m}})\left(\frac{Z}{\cos\theta_{i,m}}\right)^{2}+\frac{(\kappa(\theta_{\mathrm{i},\mathrm{m}})r_{\mathrm{v}}-1)^{2}}{4\kappa(\theta_{\mathrm{i},\mathrm{m}})r_{\mathrm{v}}}},
\end{equation*}
where $Z=U_{\mathrm{0}}/\hbar\sqrt{v_{\mathrm{F}}^{\mathrm{N}}v_{\mathrm{F}}^{\mathrm{S}}}$ is the dimensionless barrier strength introduced by Blonder and Tinkham~\cite{BlonderTinkhamPhysRevB1983}, with $U_0$ the barrier height, and $r_{\mathrm{v}}=v_{\mathrm{F}}^{\mathrm{N}}/v_{\mathrm{F}}^{\mathrm{S}}$ is the ratio of Fermi velocities in the normal conductor and in the superconductor. The term $\kappa(\theta_{\mathrm{i},\mathrm{m}})$ is given as\begin{equation*}
\kappa(\theta_{\mathrm{i},\mathrm{m}})=\cos/\sqrt{1-r_{\mathrm{k}}^{2}\sin^{2}(\theta_{\mathrm{i},\mathrm{m}})},
\end{equation*}
where $r_{\mathrm{k}}=k_{\mathrm{F}}^{\mathrm{N}}/k_{\mathrm{F}}^{\mathrm{S}}$ the ratio of wave numbers in the normal conductor and superconductor. The critical incident angle $\theta_{\mathrm{c}}=\arcsin{(1/r_{\mathrm{k}})}$ has the effect of cutting off contributions of many trajectories. The critical angle accounts for the suppression of the Andreev reflection probability when momentum cannot be conserved due to the parallel momentum in the normal region exceeding the Fermi momentum of the superconductor. For additional discussion of the modelling and simulation results under different conditions see the Supplemental Material Section \textbf{VI}. In all simulations we have assumed that $r_{\mathrm{v}}=r_{\mathrm{k}}=1$.

\vspace{0.1cm}

\section*{Acknowledgements}

We thank Herbert Kertz for technical assistance and Vladan Brajovic for help with sample characterization. Dr. Florian Lentz, Dr. Stefan Trellenkamp, and Rainer Benczek are gratefully acknowledged for their help with the required e-beam lithography and deposition of the superconducting material. We thank Andrei Manolescu, Mikio Eto, and Rui Sakano for fruitful discussions. Most of the fabrication has been performed in the Helmholtz Nano Facility at Forschungszentrum J\"ulich \cite{Albrecht2017}. This work was partly funded by Deutsche Forschungsgemeinschaft (DFG, German Research Foundation) under Germany’s Excellence Strategy—Cluster of Excellence Matter and Light for Quantum Computing (ML4Q) EXC 2004/1—390534769. This work was supported by JSPS Grants-in-Aid for Scientific Research (S) (No. 19H05610), JSPS Kakenhi (No. 19H00867) and a SPDR Fellowship provided by the "Young Scientist Program".

\section*{Author contribution}
B.\,B., C.\,K. and A.\,P. performed the nanowire growth by MBE and the deposition of the superconducting shell. F.\,B. prepared the substrates for the nanowire growth, fabricated the samples without superconducting shell and carried out the normal transport measurements. P.\,Z. and M.\,D.\,R. prepared the device chips for the DC and AC junction measurements and fabricated the Josephson junctions. P.\,Z. and R.\,S.\,D. carried out the low-temperature experiments on Josephson junctions and the subsequent data analysis. R.\,J. and A.\,M.\,S. carried out the TEM and EDX studies. P.\,Z., F.\,B., A.\,P., R.\,S.\,D. and T.\,S. wrote the manuscript. All authors contributed to the discussions.

\section*{Data availability}
All the data that support the plots and the other findings of this study are available from the corresponding author upon reasonable request.

\noindent 

%


\newpage
\clearpage
\widetext

\setcounter{section}{0}
\setcounter{equation}{0}
\setcounter{figure}{0}
\setcounter{table}{0}
\setcounter{page}{1}
\makeatletter
\renewcommand{\thesection}{S\Roman{section}}
\renewcommand{\thesubsection}{\Alph{subsection}}
\renewcommand{\theequation}{S\arabic{equation}}
\renewcommand{\thefigure}{S\arabic{figure}}
\renewcommand{\figurename}{Supplementary Figure}
\renewcommand{\bibnumfmt}[1]{[S#1]}
\renewcommand{\citenumfont}[1]{S#1}
\begin{center}
\textbf{Supplemental Material: Flux-periodic supercurrent and magnetoconductance oscillations in epitaxially-grown GaAs/InAs/Al core/shell/half-shell nanowires}
\end{center}

\section{TEM cross section images showing nanowires facets}

In Supplementary Figure~\ref{fig_Supp:Al-InAs-interface-TEM} (a) a cross section annular dark field scanning transmission electron microscope (ADF-STEM) image of a GaAs/InAs core/shell nanowire partly covered by an Al shell is shown with the detail of the Al/InAs interface given in (b). An ADF-STEM image  and an intensity profile across the GaAs/InAs/Al heterostructure are shown in Supplementary Figures \ref{fig_Supp:Al-InAs-GaAs-line-TEM} (a) and (b), respectively. The green rectangle in (a) indicates the range where the intensity profile was taken.
\begin{figure*}[!h]
	\centering
\includegraphics[width=0.98\linewidth]{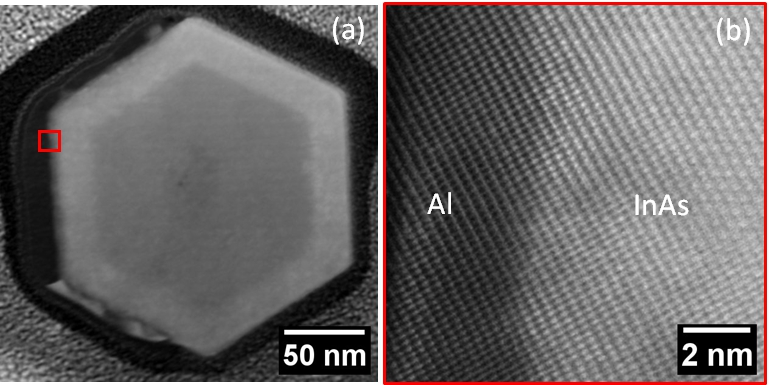}
	\caption[]{(a) Cross section ADF-STEM image of a nanowire revealing the hexagonal (b) high magnification image at the InAs/Al interface.}
	\label{fig_Supp:Al-InAs-interface-TEM}
\end{figure*}

\begin{figure*}[!h]
	\centering
\includegraphics[width=0.98\linewidth]{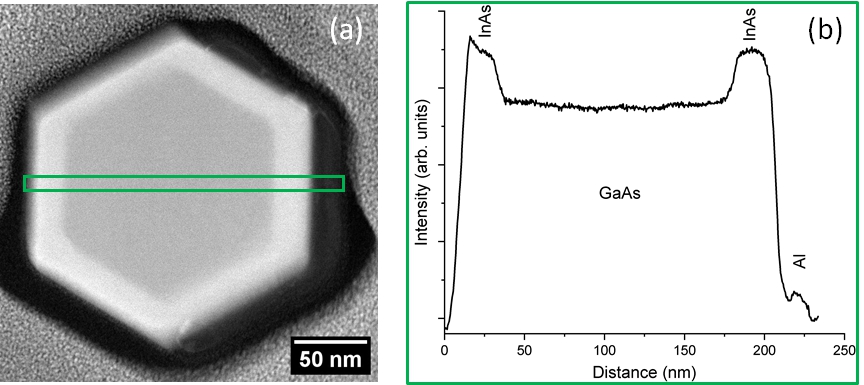}
	\caption[]{(a) ADF-STEM image of a of a nanowires and (b) intensity profile across the GaAs/InAs/Al heterostructure.}
	\label{fig_Supp:Al-InAs-GaAs-line-TEM}
\end{figure*}

\section{Nanowire diameters and shell thicknesses}

A fabricated growth substrate contains arrays of holes, with nominal hole diameters of 40, 60, and 80\,nm defined in the oxide mask. For each hole size the substrate contains pitch arrays with hole distances: 0.5, 1.0, 2.0, and 4.0\,$\upmu$m, therefore comprising different pitch types with respect to variation of these parameters. Due to different hole diameter and pitch sizes, the GaAs core diameter as well as the total diameter of the GaAs/InAs core/shell nanowire varies. The typical length of the nanowires was 4\,$\upmu$m. The results are summarized in Table~\ref{tab:diameters198} and Table~\ref{tab:diameters202}, referring to geometrical specifications for the bare GaAs/InAs and the GaAs/InAs core/shell covered by a 20-nm-thick Al half-shell  nanowire types, respectively.

\begin{table}[h!]
\centering
 \begin{tabular}{|c| c c c c|} 
 \hline
  \backslashbox{Hole size (nm)}{ \\ Hole distance ($\upmu$m) } & 0.5 & 1.0 & 2.0 & 4.0 \\ 
  \hline
 40 & 108/147 & 124/176 & -/- & 163/209\\
 60 & 104/151 & 130/171 & 130/200 & 116/188\\
 80 & 116/161 & 132/174 & 131/200 & 138/205\\ 
 \hline
 \end{tabular}
 \caption{Nanowire geometries presented with GaAs core diameter/total GaAs/InAs core/shell diameter, for pitch arrays with different hole sizes and hole distance for the bare GaAs/InAs nanowire growth. Values are given in nm unit.} \label{tab:diameters198}
\end{table}

\begin{table}[h!]
\centering
 \begin{tabular}{|c| c c c c|} 
 \hline
  \backslashbox{Hole size (nm)}{ \\ Hole distance ($\upmu$m) } & 0.5 & 1.0 & 2.0\ & 4.0 \\ 
  \hline
 40 & 117/179 & 134/210 & 139/242 & 142/259\\
 60 & 115/166 & 144/221 & 153/242 & 134/256\\
 80 & 127/173 & 146/223 & 152/235 & 151/260\\ 
 \hline
 \end{tabular}
 \caption{Nanowire diameters of GaAs/InAs core/shell nanowires comprising a superconducting Al half-shell. Labels and units refer to the same parameters as in Table~\ref{tab:diameters198}.} \label{tab:diameters202}
\end{table}


\section{Temperature-dependence of conductance in normal contacted core/shell nanowires} 

In Supplementary Figure~\ref{fig_Supp:Oszis-Temp} (a) the magnetoconductance of a normal contacted GaAs/\-InAs core/shell nanowire (device A) is shown for temperatures in the range between 1.7 and 20\,K. For this measurement an AC bias current of 20\,nA was used. The corresponding fast Fourier spectrum after subtraction of the slow vary background is depicted in Supplementary Figure~\ref{fig_Supp:Oszis-Temp} (b). With increasing temperature the oscillation amplitude decreases. This is attributed to a loss of phase coherence due to an increase of inelastic electron-electron scattering. 
\begin{figure*}[!h]
	\centering
\includegraphics[width=0.7\linewidth]{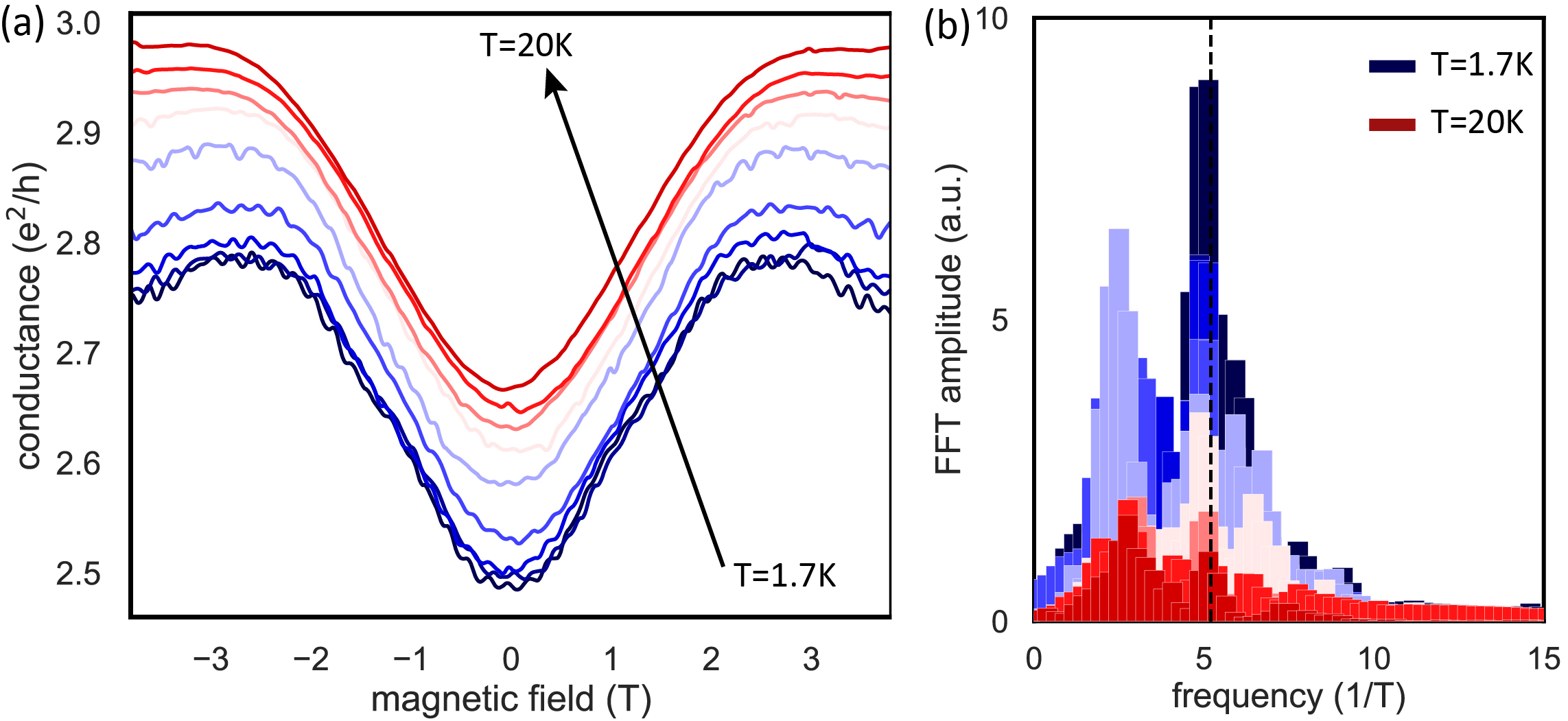}
	\caption[]{(a) Temperature-dependent magnetoconductance of device A for an AC bias current of 20\,nA at temperatures in the range between 1.7 and 20\,K. (b) Corresponding fast Fourier spectra. }
	\label{fig_Supp:Oszis-Temp}
\end{figure*}


\section{Additional data for the nanowire Josephson junction device C}

\begin{figure*}[!h]
	\centering
\includegraphics[width=0.73\linewidth]{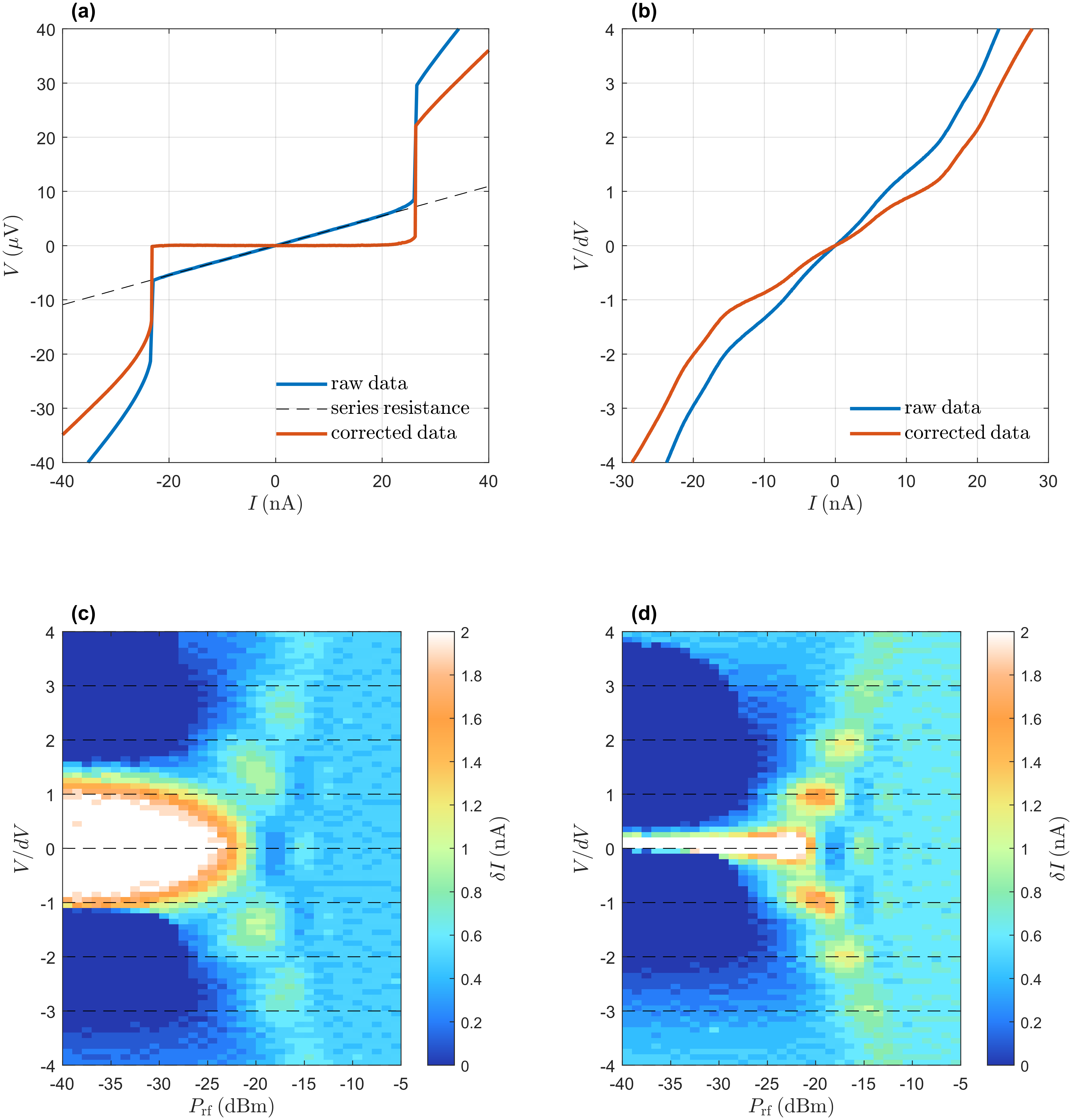}
	\caption[]{Evaluation of series resistance in device C. (a) $I-V$ trace before and after correction. The series resistance subtracted is indicated with a dashed line and is found to be consistently $R_{\mathrm{series}}\sim 273\,\mathrm{\Omega}$. (b) Demonstration of Shapiro steps plotted both as raw data and after correction with $f=2.8\,\mathrm{GHz}$ and $P_{\mathrm{rf}}=-21\,\mathrm{dBm}$. The y-axis is normalized by the expected step voltage $\Delta V=hf/2e$. (c) Histograms of raw measurement voltages plotted in a false color plot for a range of rf power ($P_{\mathrm{rf}}$). (d) Histograms of corrected measurement voltages plotted in a false color plot for a range $P_{\mathrm{rf}}$ showing good quantization of the Shapiro steps at multiples of $\Delta V$.}
	\label{fig_temp:SampleCResistance}
\end{figure*}

Device C was found to exhibit a small resistance ($R_\mathrm{series}\sim 273\,\mathrm{\Omega}$) in the supercurrent branch of the $I-V$ characteristic likely due to a poor contact to the InAs shell caused by over etching into the GaAs core during fabrication (cf. Supplementary Figure~\ref{fig_temp:SampleCResistance} (a)). The presence of this series correction to the resistance was confirmed through measurement of the Shapiro response of the junction. We find that a correction for the resistance observed in the supercurrent branch is necessary to produce well-quantized Shapiro steps, as illustrated in Supplementary Figure~\ref{fig_temp:SampleCResistance} (b). Through this method we confirm that the observed resistance is in series with the junction and not for example the effects of phase diffusion. In Supplementary Figures~\ref{fig_temp:SampleCResistance} (c) and (d) histograms of raw and corrected measurement voltages are plotted in a false color plot for a range of rf power ($P_{\mathrm{rf}}$). In the latter case a good quantization of the Shapiro steps at multiples of $\Delta V=hf/2e$ is found. We note that device B had no evidence of a series resistance and displayed clear supercurrent branches.

\begin{figure*}[!h]
	\centering
\includegraphics[width=0.7\linewidth]{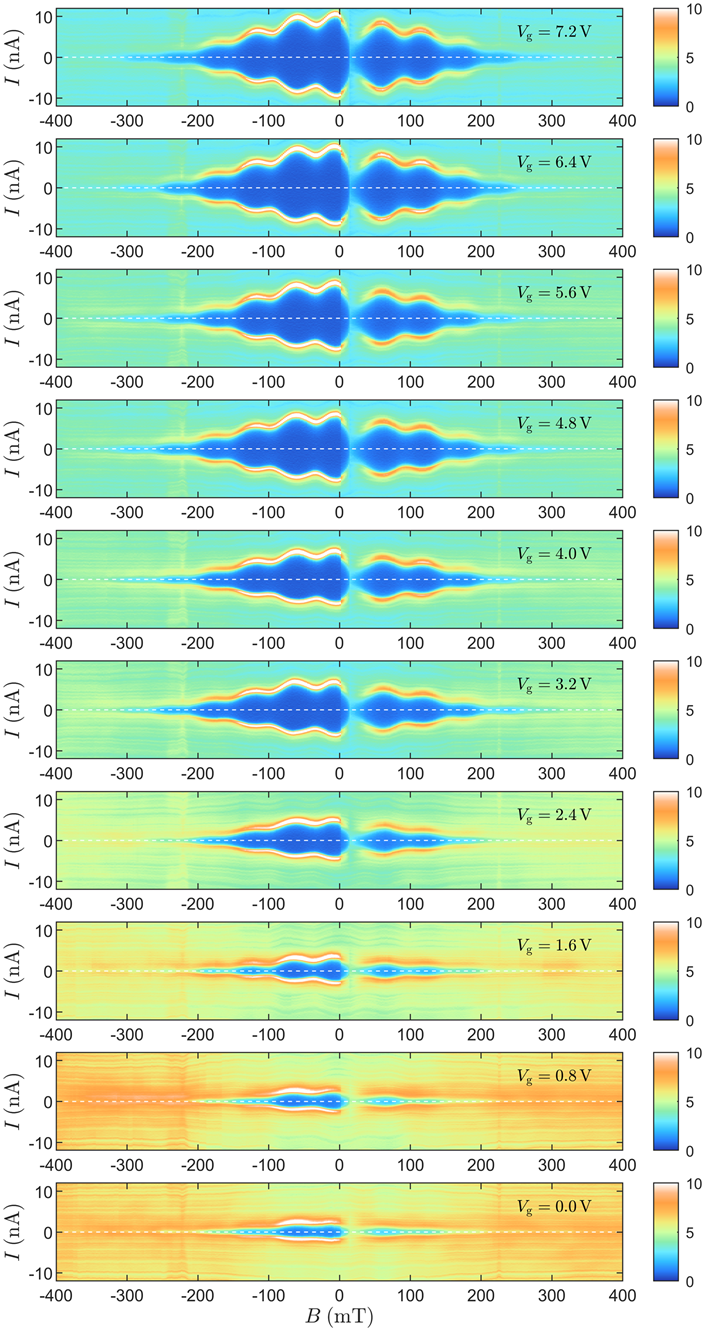}
	\caption{$dV/dI$ plots as a function of the axial magnetic field and the bias current for device C at different gate voltages. This is the experimental data used to produce Fig.~5 (a) of the main text.}
	\label{fig-temp-DevCVgdataNo1}
\end{figure*}

Supplementary Figure~\ref{fig-temp-DevCVgdataNo1} shows false colour plots of $dV/dI$ as a function of the axial magnetic field and the bias current for device C with a range of gate voltages between $V_{\mathrm{g}}=0$ and $7.2\,\mathrm{V}$. This data is analyzed to produce Fig.~5 (a) of the main text. Note that a subtraction of the series resistance, as discussed above, has been performed.

\clearpage

\section{Transport of normal contacted wire without junction}

\begin{figure*}[!h]
\centering
\includegraphics[width=0.68\linewidth]{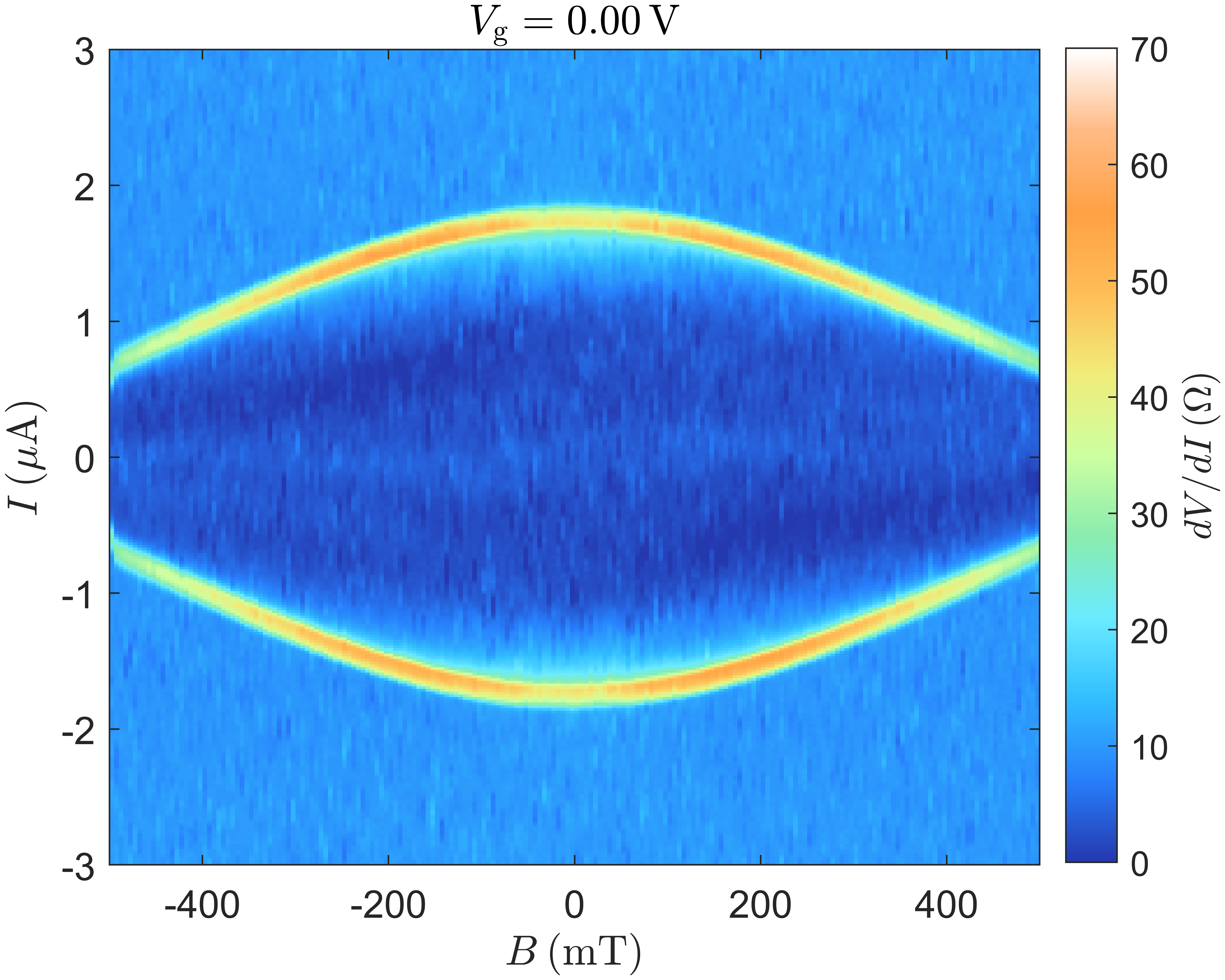}
\caption{Differential resistance $dV/dI$ as a function of magnetic field and bias current for a normal contacted Al half-shell covered GaAs/InAs core/shell nanowire device without an etched Josephson junction.}
	\label{fig_supp:NormalContactDevice}
\end{figure*}

Here, we present transport data for a nanowire with epitaxial half-shell, normal contacts and without an etched junction (cf. Supplementary Figure \ref{fig_supp:NormalContactDevice}). Contacts were formed using electron beam deposited Ti/Au with a short argon milling etch. The resulting contacts to the InAs shell are relatively poor, likely due to over etching of the thin InAs, leading to a small resistance in the supercurrent branch of the $I(V)$ of a few Ohms. A clear transport feature indicating the supercurrent is observed and exhibits no oscillatory behavior supporting the conclusions presented in the main text that the oscillations are a result of the junction transport.

\afterpage{\clearpage}

\section{Semiclassical model for Josephson junction magnetic response}

Here we provide additional details of the semi-classical model employed in the main text and described in the methods section in the main manuscript. In Supplementary Figure~\ref{Supp:SemiClassicalModelGeometry} (a) the geometry of a nanowire-based Josephson junction is illustrated with three example types of trajectories, while Supplementary Figure~\ref{Supp:SemiClassicalModelGeometry} (b) shows trajectories passing through a single point along $z_{\mathrm{cut}}$ with a single crossing of the normal region at the underside of the wire.
\begin{figure*}[h]
	\centering
    \includegraphics[width=0.98\linewidth]{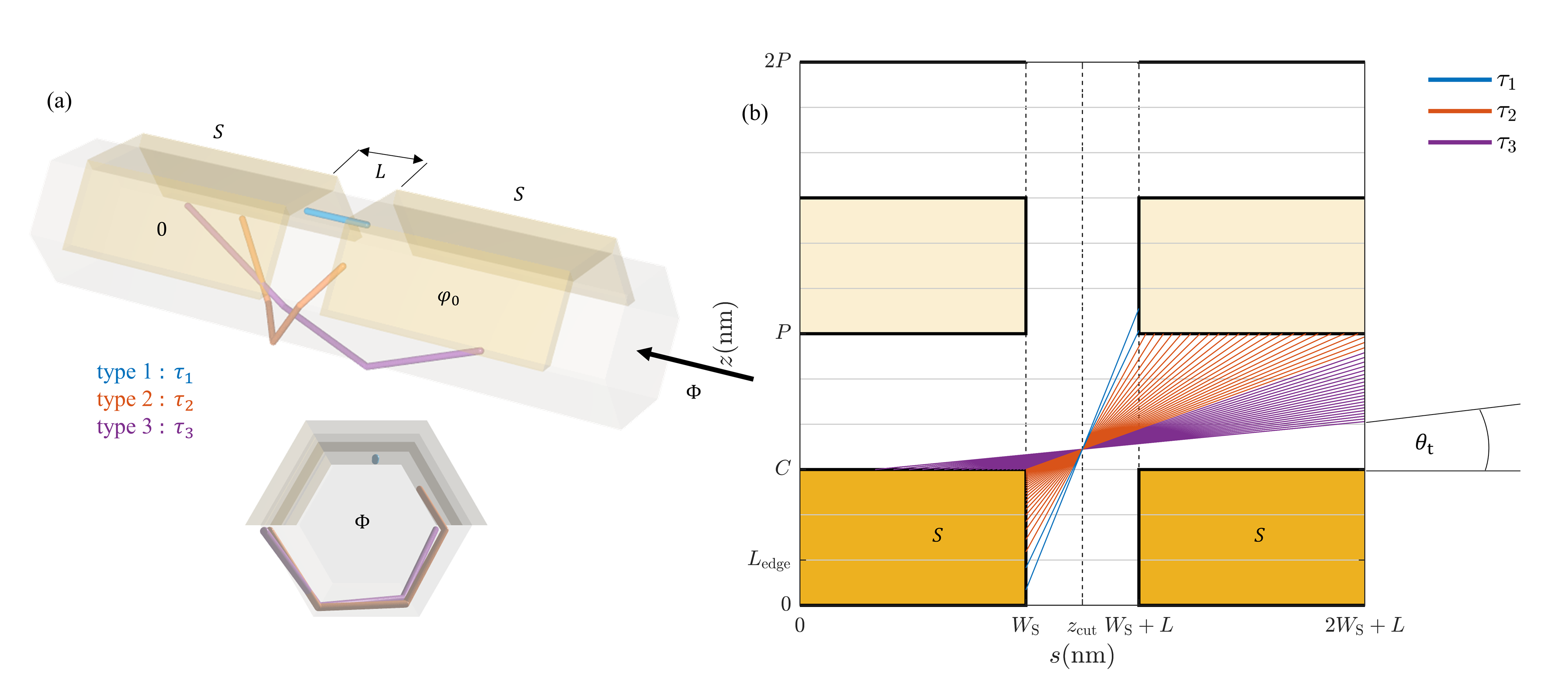}
	\caption{Simulation geometry: (a) Illustration of a nanowire Josephson junction covered by two closely spaced Al half-shell electrodes with three example types of trajectory. (b) Example trajectories passing through a single point along $z_{\mathrm{cut}}$ with a single crossing of the normal region at the underside of the wire.}
	\label{Supp:SemiClassicalModelGeometry}
\end{figure*}

To illustrate the functional form of expressions for the transparencies $\tau(\theta_{\mathrm{t}})$, with $\theta_{\mathrm{t}}$ the trajectory angle, we plot examples for $r_{\mathrm{k}}=r_{\mathrm{v}}=1$ in Supplementary Figure \ref{Supp:TauPlot}. The transparencies are subdivided in $\tau_1$, $\tau_2$, and $\tau_3$ according to the different types of trajectories defined in the main text. As can be seen, the Andreev reflection probability is rapidly suppressed for larger nanowire off-axis trajectory angles. Long paths would also be suppressed due to the mean free path of carriers in the InAs. In the simulations presented in the main text and here we limit trajectories to those with $0$ or $1$ traverse about the wire perimeter, assuming that longer trajectories would be suppressed due to the mean-free path of carriers in the InAs shell.
\begin{figure*}[h]
	\centering 
    \includegraphics[width=0.58\linewidth]{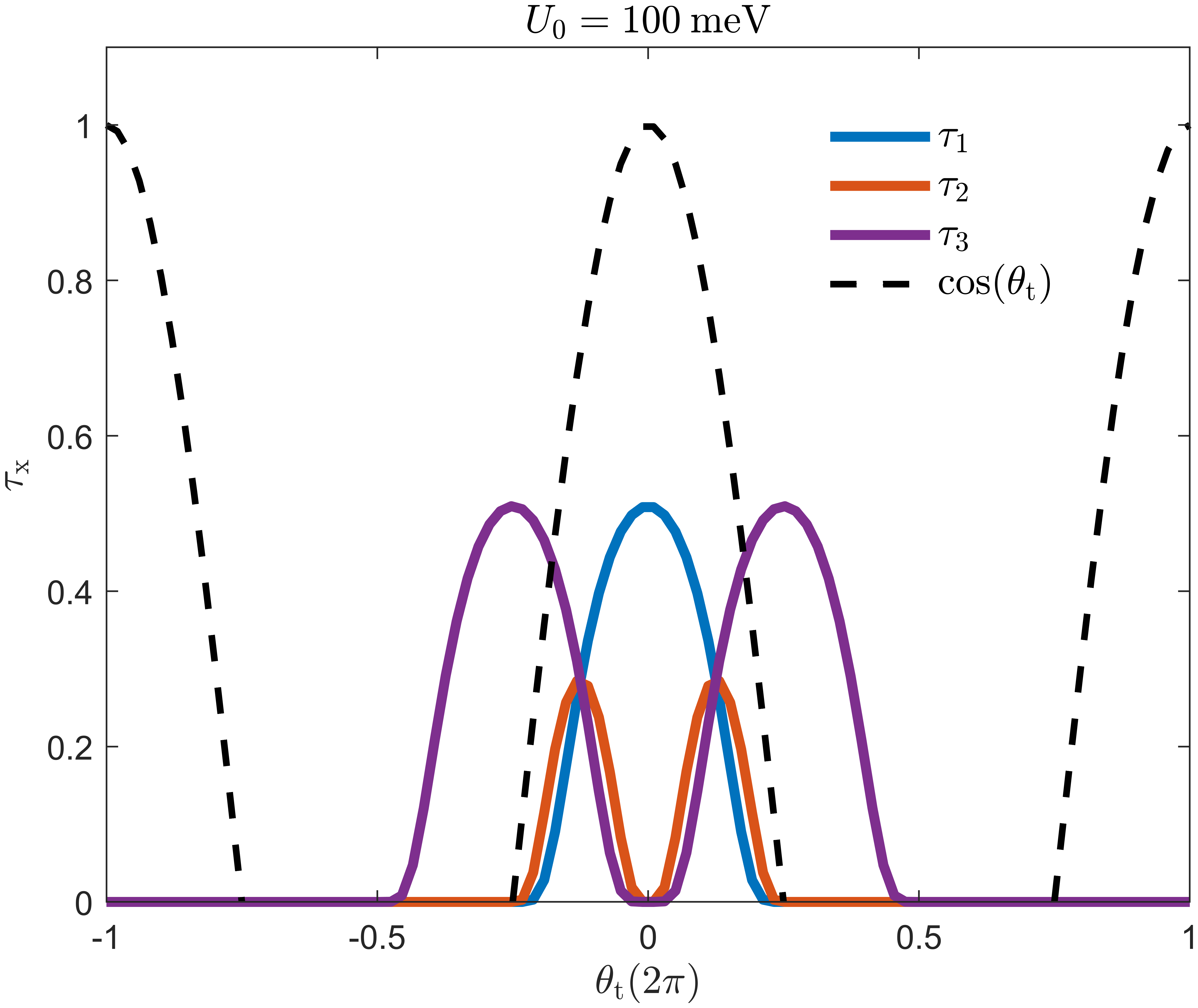}
	\caption{Transparencies $\tau$ as a function of trajectory angle $\theta_{\mathrm{t}}$ for different trajectory types.}
	\label{Supp:TauPlot}
\end{figure*}

In Supplementary Figures \ref{Supp:SemiClassicalSimA} and \ref{Supp:SemiClassicalSimB} additional simulation results are  plotted with variations of the junction length $L$ with three facets covered and the number of facets covered in superconductor for a fixed length of 100\,nm, respectively. Here, we see that the modification of the available trajectories can have an effect on visibility of the features. Our simulations do not account for the scattering in the InAs shell assuming that all trajectories are ballistic. However, in reality at longer channel lengths diffusive transport would be dominant and likely rapidly suppress the visibility of the features presented. A theoretical treatment of the problem in this regime would be of interest but is beyond the scope of the present study.
\begin{figure*}[h]
	\centering 
    \includegraphics[width=0.58\linewidth]{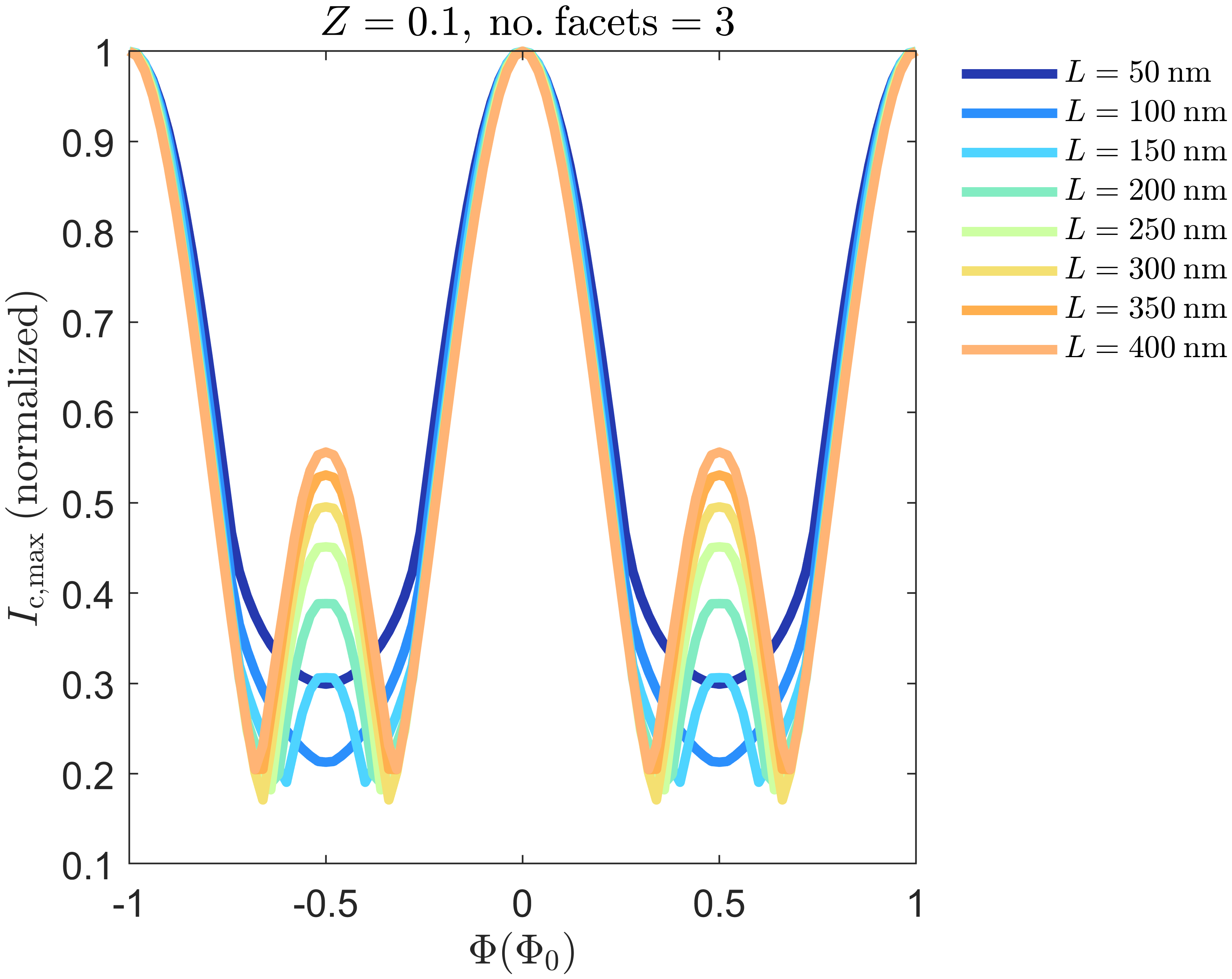}
	\caption{Simulations of the critical current as a function of normalized magnetic flux $\Phi/\Phi_0$ with $Z=0.1$, three nanowire facets covered by the superconducting contacts and a range of junction length $L$ between 50 and 400\,nm.}
	\label{Supp:SemiClassicalSimA}
\end{figure*}

\begin{figure*}[h]
	\centering 
	\includegraphics[width=0.58\linewidth]{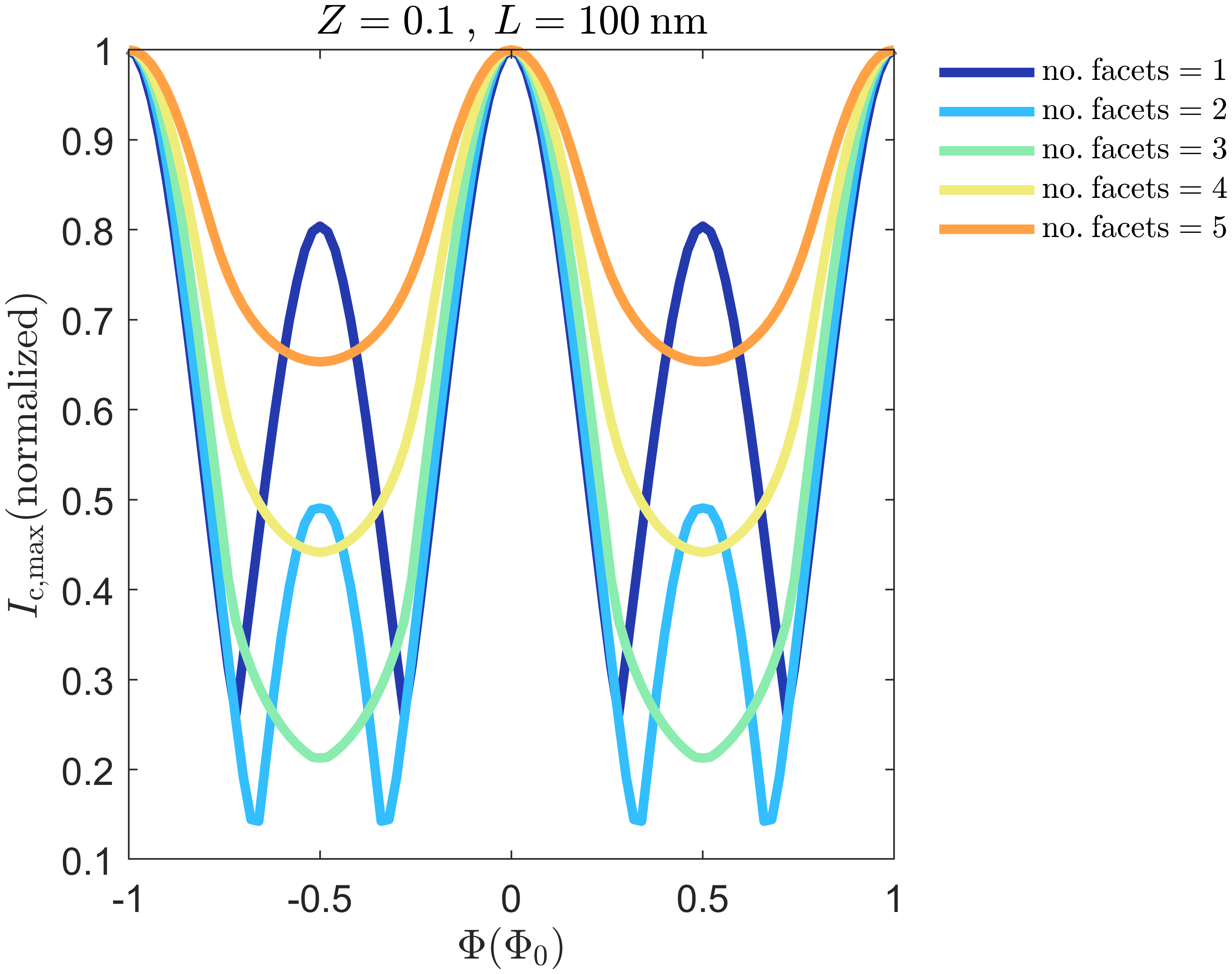}
 \caption{Simulations of the critical current as a function of normalized magnetic flux $\Phi/\Phi_0$ with $Z=0.1$, $L=100\,\mathrm{nm}$ and different number of nanowire facets covered by the superconducting contacts.}
	\label{Supp:SemiClassicalSimB}
\end{figure*}

\clearpage

\section{Comparison of angular response with full-shell nanowire devices}

We note that the observed apparent reduction of oscillation period $\Delta B$ of the differential resistance $dV/dI$ with off-axis application of magnetic field is similar to an effect observed in InAs nanowire devices with a full aluminium shell. As an example, in Supplementary Figure~\ref{fig_temp:FullShellNWJJ} we show experimental data showing Little-Parks oscillations in an Al full-shell InAs nanowire. Data here is collected using a lock-in measurement technique with a relatively large AC current excitation, a technique at the time used to quickly check or the alignment of magnetic field and nanowire axis. The observed resistance oscillations of the differential resistance as a function of field angle and magnetic field amplitude shown in Supplementary Figure~\ref{fig_temp:FullShellNWJJ} (a) indicate the region with non-dissipative superconducting transport and the features of destructive Little-Parks oscillations \cite{LittleParks1962,Vaitieke2020,VaitiekenasScience2020,Vekris2021}. The experimentally determined oscillation period $\Delta B$ as a function of tilt angle $\theta$ is given in Supplementary Figure~\ref{fig_temp:FullShellNWJJ} (b). For comparison, the dashed red curve in the graph corresponds to the expected oscillation period assuming that only the axial component of the magnetic field produces the oscillations, following a cosine dependence typical for conventional Aharonov-Bohm-type oscillations, universal conductance fluctuations or weak localization \cite{Haas2016,Jespersen2015,Dong2010}. We also considered the expected periodicity if the oscillation period was fixed via the flux penetrating the elliptical cross section of the nanowire (solid blue line). Both calculated curve do not match to the observed dependence of $\Delta B$.  
\begin{figure*}[!h]
	\centering
\includegraphics[width=0.9\linewidth]{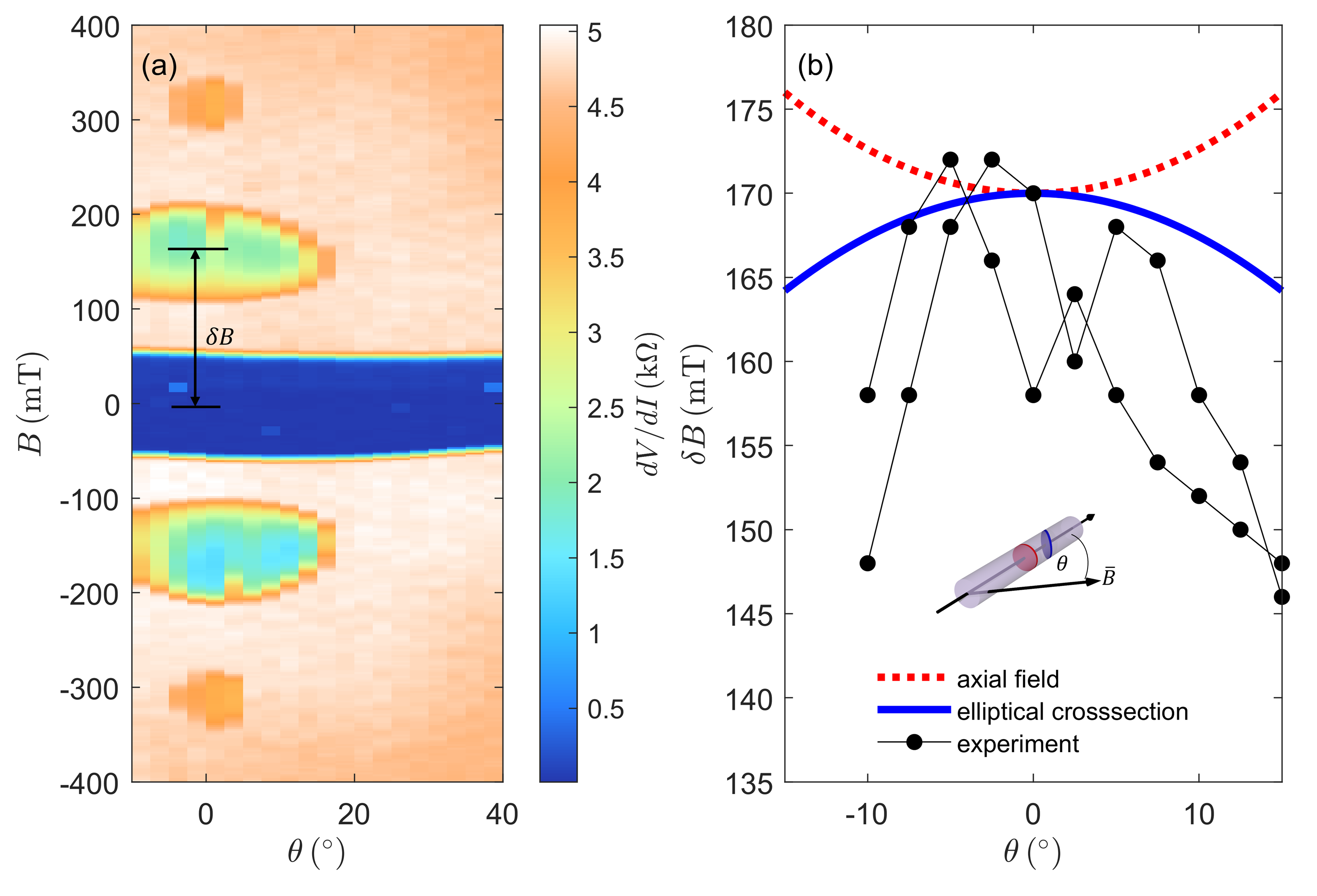}
\caption{(a) Field angle dependence of the differential resistance as a function of the magnetic field $B$ for a full-shell InAs nanowire Josephson junction device. A gate voltage of  $V_{\mathrm{g}}=6.0\,\mathrm{V}$ is applied to ensure that the device channel is conducting and that the critical current is maximised. (b) Experimentally determined oscillation period $\Delta B$ (black dots) in comparison to the theoretically expected period taking an elliptical cross section into account (solid blue line) or restricting to the axial component of the magnetic field only (red dashed line).}
	\label{fig_temp:FullShellNWJJ}
\end{figure*}

In the case of the response of full-shell nanowires to an off-axis magnetic field, similar behavior can be replicated using a theoretical model introduced by Vekris \textit{et al.} in Ref.~\cite{Vekris2021}. For a full description of the model we direct readers to Ref.~\cite{Vekris2021}, as the implementation here is identical. Briefly, the model assumes a single hollow thin-walled aluminium cylinder with diameter $d_{\mathrm{F}}$ and wall thickness $t_{\mathrm{s}}$. The critical superconducting temperature $T_{\mathrm{c}}(B)$ is given as
\begin{equation*}
    \ln\left(\frac{T_{\mathrm{c}}(\alpha)}{T_{\mathrm{c0}}}\right)=\Psi\left(\frac{1}{2}\right)-\Psi\left(\frac{1}{2}+\frac{\alpha}{2\pi T_{\mathrm{c}}(\alpha)}\right),
\end{equation*}
where $\Psi(z)$ is the digamma function \cite{Abrisokov1960} and $T_{\mathrm{c0}}=T_{\mathrm{c}}(B=0)$ is the critical temperature at zero temperature. The parameter $\alpha$ accounts for Cooper pair breaking \cite{Sternfeld2011,Schwiete2009,Dao2009,tinkham2004Book,Rogachev2005} and is separated into two terms containing effects of the parallel field threading the cylinder and the field perpendicular to the cylinder, $\alpha=\alpha_{\parallel}(B_{\parallel})+\alpha_{\perp}(B_{\perp})$. The Little-Parks oscillations arise from $\alpha_{\parallel}(B_{\perp})$
\begin{equation*}
    \alpha_{\parallel}(B_{\parallel})=\frac{4\xi^{2}T_{\mathrm{c0}}}{A_{\parallel}}\left[\left(n-\frac{\Phi_{\parallel}(B_{\parallel})}{\Phi_{\mathrm{0}}}\right)^{2}+\frac{t_{\mathrm{s}}^{2}}{d_{\mathrm{F}}^{2}}\left(\frac{\Phi_{\parallel}(B_{\parallel})^{2}}{\Phi_{\mathrm{0}}^{2}}+\frac{n^{2}}{3}\right)\right] \; ,
\end{equation*}
where $\xi$ is the coherence length, $\Phi_{\parallel}(B_{\parallel})=B_{\parallel}A_{\parallel}$ is the magnetic flux threading the cylinder where $A_{\parallel}=\pi d^{2}_{\mathrm{F}}/4$ is the cross sectional area, and $n$ is the number of flux quanta. The term $\alpha_{\perp}(B_{\perp})$ applies a pair breaking effect for field perpendicular to the cylinder \cite{Shah2007} given as
\begin{equation*}
    \alpha_{\perp}(B_{\perp})=\frac{4\xi^{2}T_{\mathrm{c0}}}{A_{\perp}}\frac{\Phi_{\perp}(B_{\perp})^{2}}{\Phi_{\mathrm{0}}^{2}},
\end{equation*}
where $\Phi_{\perp}(B_{\perp})=B_{\perp}A_{\perp}$. When implemented by Vekris \textit{et al.} \cite{Vekris2021} the parameter $A_{\perp}$ is used as a free fitting parameter to existing experimental data. Here, we plot a representative set of simulation results for different values of $A_{\perp}$ and show that the term can have significant effect on the appearance of the Little-Parks oscillations for misaligned fields recreating the anomalous reduced periodicity reported above. The critical current is then given as \cite{Bardeen1962}
\begin{equation*}
    I_{\mathrm{c}}(\alpha)=I_{\mathrm{c0}}\left(\frac{T_{\mathrm{c}}(\alpha)}{T_{\mathrm{c0}}}\right)^{3/2} \, .
\end{equation*}

To illustrate the effect of depairing due to an applied perpendicular magnetic field, example simulation results are shown in Supplementary Figures~\ref{fig_temp:LPtest500}, \ref{fig_temp:LPtest5000}, and \ref{fig_temp:LPtest50000}. Variation of the parameter $A_{\mathrm{\perp}}$ can control the amount of the depairing when an off-axis field is applied, thus modifying the rate of reduction of $I_{\mathrm{c}}$ as $B$ is increased. As can be seen in the Supplementary Figures \ref{fig_temp:LPtest500} or \ref{fig_temp:LPtest5000},  relatively small $A_{\mathrm{\perp}}$ results in destructive Little-Parks effects with a period determined by the threading flux. However when $A_{\mathrm{\perp}}$ is large, as in Supplementary Figure \ref{fig_temp:LPtest50000}, the faster suppression of $I_{\mathrm{c}}$ causes the appearance of a decreasing oscillation period as observed in our experimental result above. These results suggest that the depairing effect due to the off-axis field may also account for the apparent shortening of the oscillation period shown in the main text and that this effect can be viewed as a consequence of a faster reduction of critical current in the presence of an increasing perpendicular magnetic field.
\begin{figure*}[p]
	\centering
\includegraphics[width=0.7\linewidth]{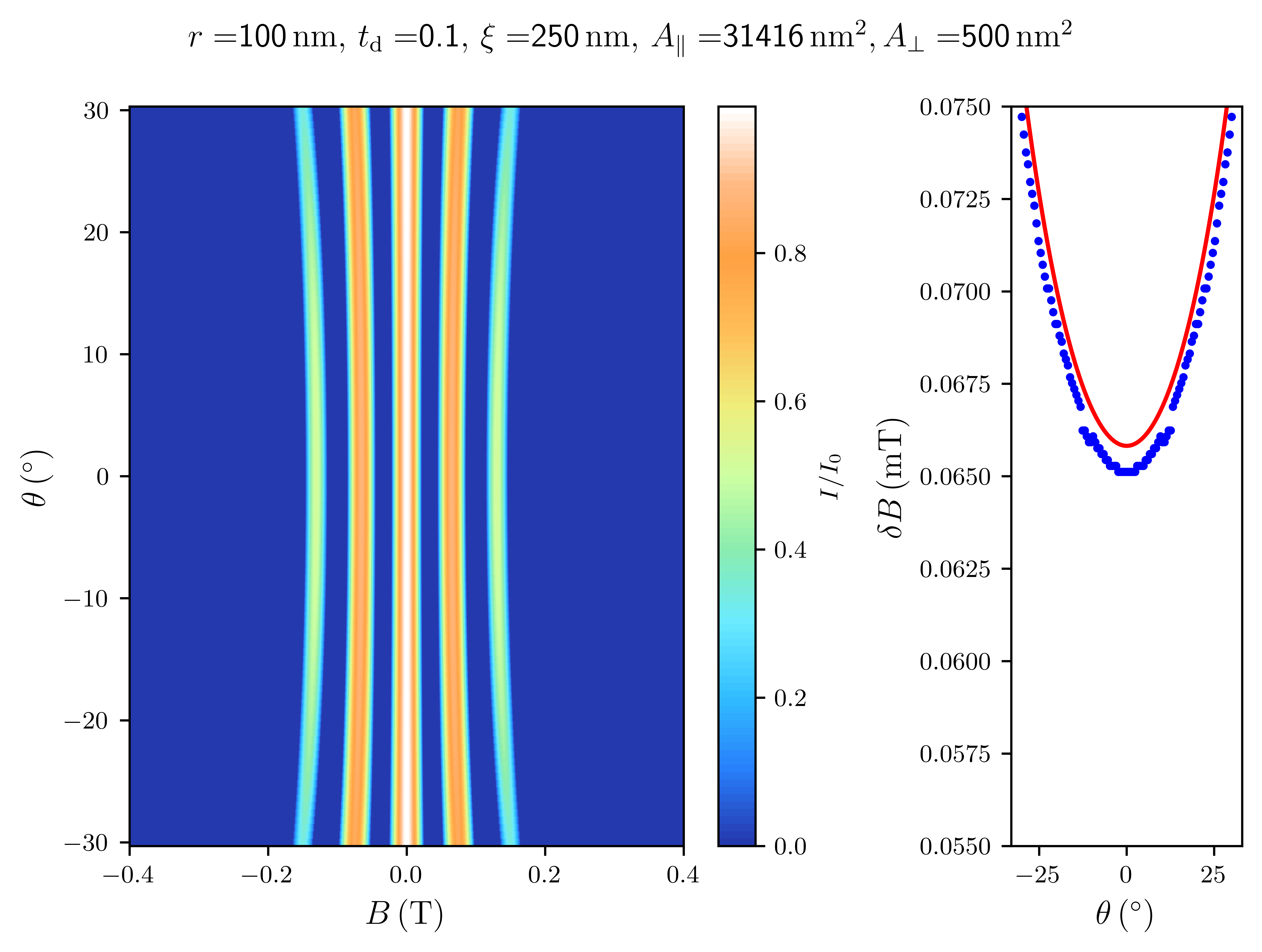}
	\caption{Simulation of the Little-Parks effect: (left panel) Normalized critical current $I/I_0$, for a superconducting cylinder with a magnetic field $B$ applied at an angle $\theta$ to the cylinder axis and $A_{\mathrm{\perp}}=500\,\mathrm{nm^{2}}$. Other parameters are as indicated in the figure. (right panel) The oscillation period $\Delta B$ is extracted as the separation of the $B=0$ peak and first peak at positive $B$ and compared with the expected oscillation period considering only the flux threading the cylinder (red line).}
	\label{fig_temp:LPtest500}
\end{figure*}

\begin{figure*}[p]
	\centering
\includegraphics[width=0.7\linewidth]{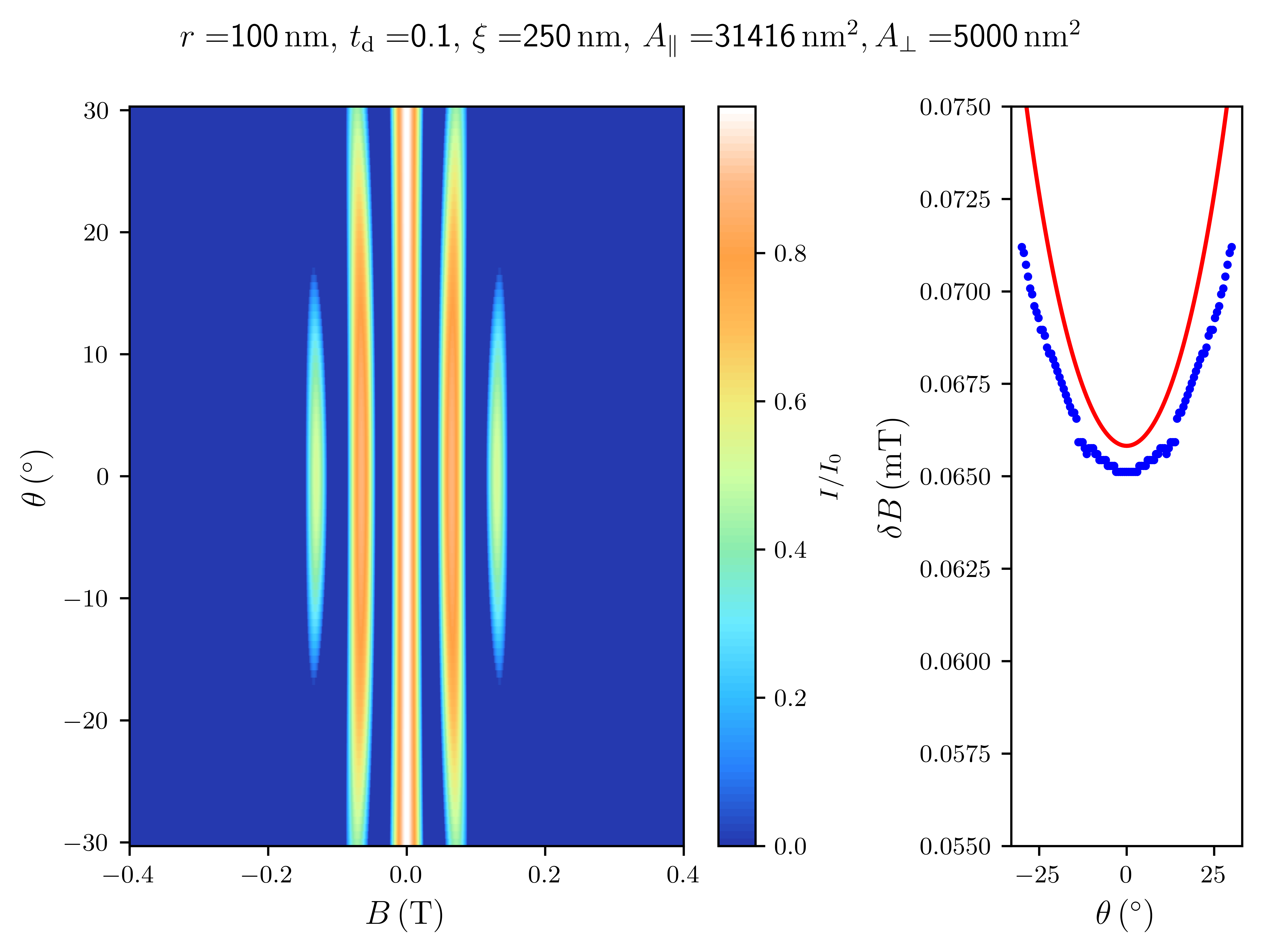}
	\caption{Simulation of Little-Parks effect: (left panel) Normalized critical current $I/I_0$, for a superconducting cylinder with a magnetic field $B$ applied at an angle $\theta$ to the cylinder axis and $A_{\mathrm{\perp}}=5000\,\mathrm{nm^{2}}$. Other parameters are as indicated in the figure. (right panel) The oscillation period $\Delta B$ is extracted as the separation of the $B=0$ peak and first peak at positive $B$ and compared with the expected oscillation period considering only the flux threading the cylinder (red line).}
	\label{fig_temp:LPtest5000}
\end{figure*}

\begin{figure*}[t]
	\centering
\includegraphics[width=0.7\linewidth]{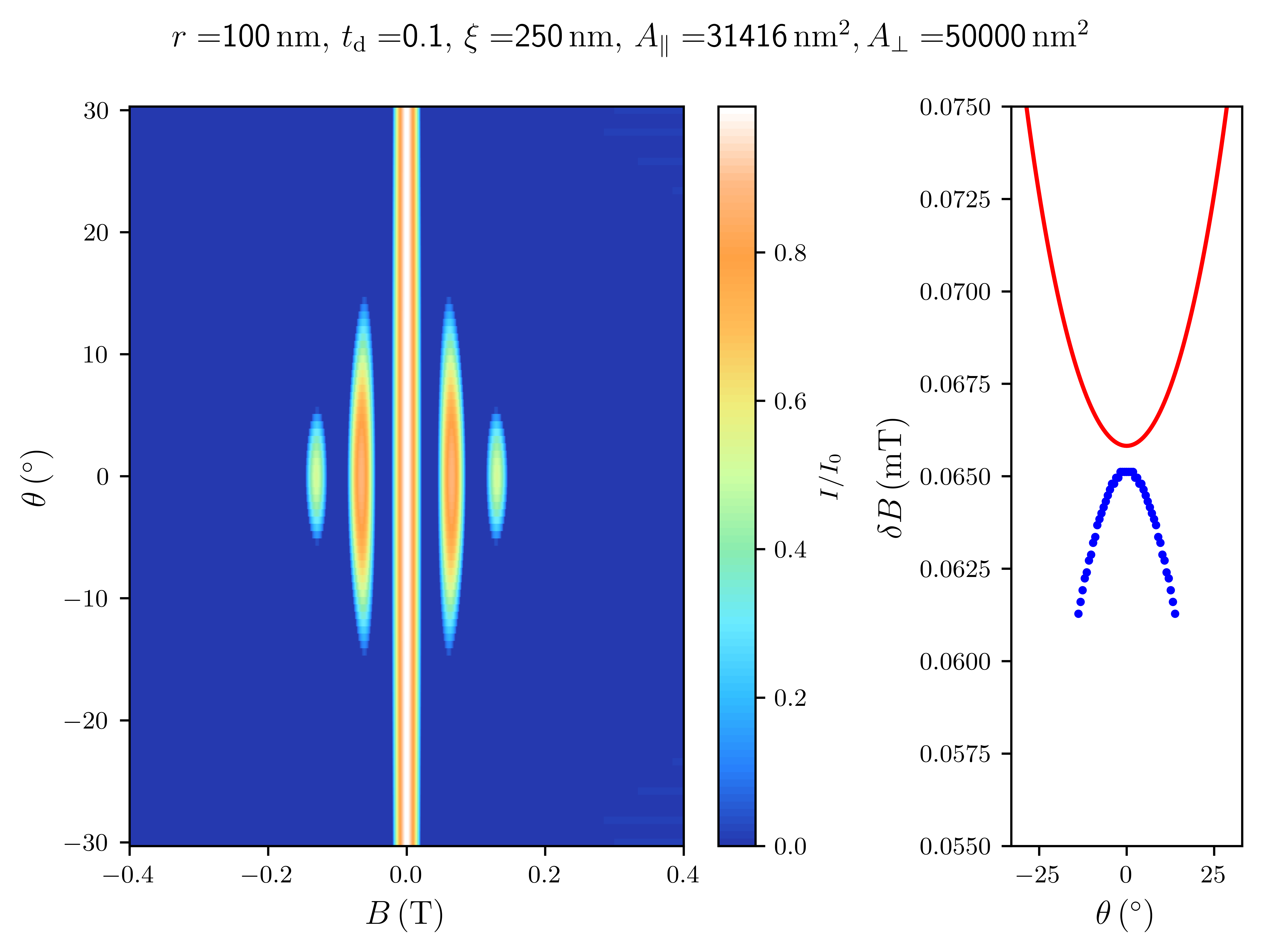}
	\caption{Simulation of Little-Parks effect: (left panel) Normalized critical current $I/I_0$, for a superconducting cylinder with a magnetic field $B$ applied at an angle $\theta$ to the cylinder axis and $A_{\mathrm{\perp}}=50000\,\mathrm{nm^{2}}$. Other parameters are as indicated in the figure. (right panel) The oscillation period $\Delta B$ is extracted as the separation of the $B=0$ peak and first peak at positive $B$ and compared with the expected oscillation period considering only the flux threading the cylinder (red line).}
	\label{fig_temp:LPtest50000}
\end{figure*}

\newpage
\newpage

%

\end{document}